\documentclass[journal]{IEEEtran}
\usepackage{amsmath,amssymb,amsfonts}
\usepackage{bm}
\usepackage{array}
\usepackage{color}
\usepackage{textcomp}
\usepackage{stfloats}
\usepackage{url}
\usepackage{verbatim}
\usepackage{graphicx}
\usepackage{subfigure}
\usepackage{cite}
\usepackage{epstopdf}
\usepackage{booktabs}
\allowdisplaybreaks[4]
\usepackage[linesnumbered,boxed,ruled,commentsnumbered]{algorithm2e}
\usepackage{amsthm}
\usepackage{cuted}
\theoremstyle{remark}
\newtheorem{theorem}{Theorem}
\newtheorem{lemma}{Lemma}
\newtheorem{corollary}{Corollary}

\begin{document}

\title{Beam Structured Turbo Receiver for HF Skywave Massive MIMO}

\author{
   Linfeng Song, \IEEEmembership{Member, IEEE}, Ding Shi, \IEEEmembership{Member, IEEE}, Xiqi Gao, \IEEEmembership{Fellow, IEEE}, Geoffrey Ye Li, \IEEEmembership{Fellow, IEEE}, and Xiang-Gen Xia, \IEEEmembership{Fellow, IEEE}
	
\thanks{

	Linfeng Song, Ding Shi, and Xiqi Gao are with the National Mobile Communications Research Laboratory, Southeast University, Nanjing 210096, China, and also with Purple Mountain Laboratories, Nanjing 211111, China (e-mail: songlf@seu.edu.cn; shiding@seu.edu.cn;  xqgao@seu.edu.cn).

   Geoffrey Ye Li is with the Department of Electrical and Electronic
   Engineering, Imperial College London, London SW7 2AZ, U.K. (e-mail:
   geoffrey.li@imperial.ac.uk).

   Xiang-Gen Xia is with the Department of Electrical and Computer
   Engineering, University of Delaware, Newark, DE 19716, USA (e-mail:
   xxia@ee.udel.edu).

}

}

\maketitle

\begin{abstract}
In this paper, we investigate receiver design for high frequency (HF) skywave massive multiple-input multiple-output (MIMO) communications. We first establish a modified beam based channel model (BBCM) by performing uniform sampling for directional cosine with deterministic sampling interval, where the beam matrix is constructed using a phase-shifted discrete Fourier transform (DFT) matrix. Based on the modified BBCM, we propose a beam structured turbo receiver (BSTR) involving low-dimensional beam domain signal detection for grouped user terminals (UTs), which is proved to be asymptotically optimal in terms of minimizing mean-squared error (MSE). Moreover, we extend it to windowed BSTR by introducing a windowing approach for interference suppression and complexity reduction, and propose a well-designed energy-focusing window. We also present an efficient implementation of the windowed BSTR by exploiting the structure properties of the beam matrix and the beam domain channel sparsity. Simulation results validate the superior performance of the proposed receivers but with remarkably low complexity.
\end{abstract}
\begin{IEEEkeywords}
   Massive MIMO, HF skywave communications, beam structured turbo receiver, windowing.
\end{IEEEkeywords}
\IEEEpeerreviewmaketitle

\section{Introduction}

As the demand for global coverage continues to rise~\cite{9397776}, high frequency (HF) communications, whose frequency range is $3$ to $30$ MHz, have emerged as a highly promising technology, allowing for long-range over-the-horizon communication through ionosphere refraction~\cite{johnson2013third}. However, due to limited spectrum resource and complex ionosphere conditions, the data rates are relatively low in conventional point-to-point HF communications. Several studies have investigated the application of multiple-input multiple-output (MIMO) technology to HF communications~\cite{7775853, 7604101}, where certain performance improvement can be obtained.   

To significantly enhance the spectrum and energy efficiencies, massive MIMO, a key technology for 5G system and its evolution~\cite{5595728}, has been recently introduced in HF skywave communications~\cite{yu2021hf, HFmimo, 10636972, 10419346, BDRP, 10122719, BSD, 9819435}. The HF skywave massive MIMO system has been first proposed in~\cite{yu2021hf}, demonstrating that massive MIMO can offer remarkable performance gain for HF skywave communications. To obtain the performance gain of massive MIMO, some related techniques, such as channel acquisition~\cite{HFmimo, 10636972}, downlink precoding transmission~\cite{10419346, BDRP}, uplink signal detection~\cite{10122719, BSD} and cooperative multistation secure transmission~\cite{9819435} have been investigated.

In order to fully leverage the benefits of the HF skywave massive MIMO, there is a considerable demand for receiver design with prominent performance yet relatively low computational cost. However, the maximum likelihood (ML) receiver~\cite{1237128}, and some near-optimal nonlinear receivers (e.g., sphere decoding~\cite{7801954})
suffer from prohibitive complexity. To address this issue, Bayesian inference approaches, such as expectation propagation (EP)~\cite{6841617, 9475520}, approximate message passing (AMP)~\cite{6778065} and information geometry approach (IGA)~\cite{10416397} have been developed. Some linear receivers, such as zero-forcing (ZF) receiver and minimum mean-squared error (MMSE) receiver, are particularly popular~\cite{7244171, 8804165, 9685271}. To avoid high-dimensional matrix operations of linear receivers, some approximate methods are proposed, including Newton iteration~\cite{7370771}, polynomial expansion~\cite{7349180}, Gauss–Seidel~\cite{6954512}, recursive conjugate-gradient~\cite{8950336}, etc.

For spatially correlated massive MIMO channels, beam domain sparsity has been employed to design low-complexity receivers. A beam domain local MMSE detector is proposed for dimensionality reduction~\cite{8815585}, and enhanced in~\cite{9838343} by combining the information from the overlapped local MMSE filters in the log-likelihood ratio (LLR) domain. In~\cite{8977498}, ZF detection is implemented after channel matrix compression via beam selection. Additionally, a layered belief propagation (BP) detector is developed by integrating the sparse beam domain channel and the space domain channel~\cite{9484686}. On account of the small angle spread in HF communications, by exploiting the beam domain sparsity, there are several related works about HF skywave massive MIMO receiver design. In~\cite{10122719}, Slepian transform based detectors are derived by utilizing the index set for limited non-zero beams of each user terminal (UT). Space domain and space-frequency domain beam structured detectors are proposed in~\cite{BSD}, involving low-dimensional beam domain detector designs. Similar ideas are also applied to channel estimation~\cite{10636972} and downlink precoding~\cite{10419346, BDRP}. However, the performance of the mentioned receivers for HF skywave massive MIMO is unsatisfactory when the inter-UT interference becomes severe.

Expoiting the same spirit as turbo codes~\cite{539767}, turbo receiver has been widely adopted to improve the system performance by iteratively exchanging soft information about the coded bits between a soft-input soft-output (SISO) detector and a SISO decoder~\cite{douillard1995iterative,984761,5695128}. In the past few decades, turbo receiver has been incorporated into various communication systems, such as code division multiple access (CDMA)~\cite{774855}, space-time coded transmission~\cite{1687707}, continuous phase modulation~\cite{9187629}, and MIMO transmission~\cite{zhong2014mmse, 6778065, 10272020}. Also, turbo receiver has been introduced in vector OFDM~\cite{1658439} or orthogonal time frequency space (OTFS) modulation systems~\cite{9492800, 10018250}, exploiting the inherent sparsity of the delay-Doppler domain channel for complexity reduction. 

In this paper, we investigate the receiver design for HF skywave massive MIMO communications using the sparsity of the beam domain channel. We propose the beam structured turbo receiver for performance improvement but with low computational complexity. The main contributions are summarized as follows. 
\begin{itemize}
   \item We propose a modified beam based channel model (BBCM) for HF skywave massive MIMO using sampled steering vectors with deterministic sampling interval for directional cosine. The beam matrix composed by sampled steering vectors exhibits phase-shifted partial discrete Fourier transform (DFT) structure, which is beneficial for low-complexity receiver design without sacrificing the accuracy of channel representation.
   \item Building on the modified BBCM, we derive a beam structured turbo receiver (BSTR) by taking advantage of the beam domain channel sparsity. Specifically, we categorize all UTs into finite groups and low-dimensional beam domain signal detection for each group is designed with the MMSE criterion using the a priori information, where the used beams are the union of the non-zero beam sets of the UTs within each group. Moreover, it is proved that when the number of BS antennas is sufficiently large, the beam structured signal detection is asymptotically optimal in the sense of minimizing MSE.
   \item To further mitigate the inter-UT interference and reduce the computational complexity of the BSTR, we generalize the BSTR to windowed version by applying a window function to the received signal before signal detection. In order to effectively suppress the interference, an energy-focusing window is designed to restrain the energy diffusion of the beam domain channel. Finally, we present the efficient implementation of windowed BSTR by employing structure properties of the beam matrix and window functions.
\end{itemize}

The rest of the paper is organized as follows. In Section~\ref{sec_chan_mod}, we introduce the modified BBCM for HF skywave massive MIMO systems. In Section~\ref{BSTR}, the BSTR is proposed by taking advantage of the beam
domain channel sparsity. In Section~\ref{WBSTR}, we propose the windowed BSTR with an energy-focusing window and its efficient implementation. Simulation results are provided in Section~\ref{Sim} and the work is concluded in Section~\ref{Con}.

\emph{Notations}: Boldface lower (upper) case letters represent vectors (matrices). $\mathbb{B}$, $\mathbb{Z}$, $\mathbb{R}$ and $\mathbb{C}$ denote the sets of all binary numbers, integers, real numbers and complexity numbers, respectively. Set $\mathcal{Z}_{N}^{+}\triangleq\{1,\cdots,N\}$. Notations $(\cdot)^{\ast}$, $(\cdot)^{\mathrm{T}}$, $(\cdot)^{\mathrm{H}}$, $(\cdot)^{\dagger}$, and $\operatorname{tr}\{\cdot\}$ represent the conjugate, transpose, conjugate-transpose, Moore-Penrose inverse, and trace operations, respectively. $\mathbf{O}$ and $\mathbf{I}_{M}$ indicate zero matrix and $M$-dimensional identity matrix, respectively. $\mathbf{0}_{M}$ and $\mathbf{e}_{m}$ represent $M$-dimensional all-zero vector and the $m$-th column of identity matrix, respectively. $\operatorname{diag}\{\mathbf{a}\}$, $[\mathbf{a}]_{i}$ and $[\mathbf{A}]_{i,j}$ denote the diagonal matrix with $\mathbf{a}$ along its main diagonal, the $i$-th element of $\mathbf{a}$ and the $(i, j)$-th entry of $\mathbf{A}$, respectively. $\overline{\jmath} = \sqrt{-1}$, $\odot$, and $\otimes$ denote the imaginary unit, the Hadamard product operator, and the Kronecker product operator, respectively. $|\cdot|$ denotes the cardinality for a set or the absolute value for a scalar. $\mathbb{E}\{\cdot\}$ and $\operatorname{Cov}\{\cdot\}$ represent the ensemble expectation and covariance, respectively. $\langle \cdot\rangle_{N}$ and $\|\cdot\|_{2}$ denote the modulo-$N$ operation and the $\ell_{2}$-norm, respectively. $\lfloor \cdot\rfloor$, $\lceil \cdot\rceil$ and $\lceil\cdot\rfloor$ indicate the floor, ceiling and rounding operations, respectively. The circular symmetric complex Gaussian distribution with mean $\mathbf{a}$ and covariance $\mathbf{A}$ is denoted as $\mathcal{CN}(\mathbf{a}, \mathbf{A})$. The forward cyclic shift matrix and its variant are respectively denoted by $\bm{\Pi}_{S}\triangleq [\mathbf{e}_{2}, \mathbf{e}_{3}, \cdots, \mathbf{e}_{S}, \mathbf{e}_{1}]\in\mathbb{R}^{S\times S}$ and $\overline{\bm{\Pi}}_{S}\triangleq [\mathbf{e}_{2}, \mathbf{e}_{3}, \cdots, \mathbf{e}_{S}, -\mathbf{e}_{1}]\in\mathbb{R}^{S\times S}$. And $\mathbf{S}(N,\mathcal{T})\triangleq [\mathbf{e}_{n_{1}}, \mathbf{e}_{n_{2}}, \cdots, \mathbf{e}_{n_{T}}]\in\mathbb{R}^{N\times T}$ is the selection matrix with $\mathcal{T} = \{n_{1},\cdots,n_{T}\} \subseteq\mathcal{Z}_{N}^{+}$, and $n_{1}<\cdots<n_{T}$, $T = |\mathcal{T}|$.

\section{System Model}\label{sec_chan_mod}
In this section, we present the system configuration for HF skywave massive MIMO and introduce the modified BBCM, which is vital for subsequent receiver designs.

\subsection{System Configuration}\label{2-A}

\begin{figure}[ht]
	\centering
	\includegraphics[width=0.48\textwidth]{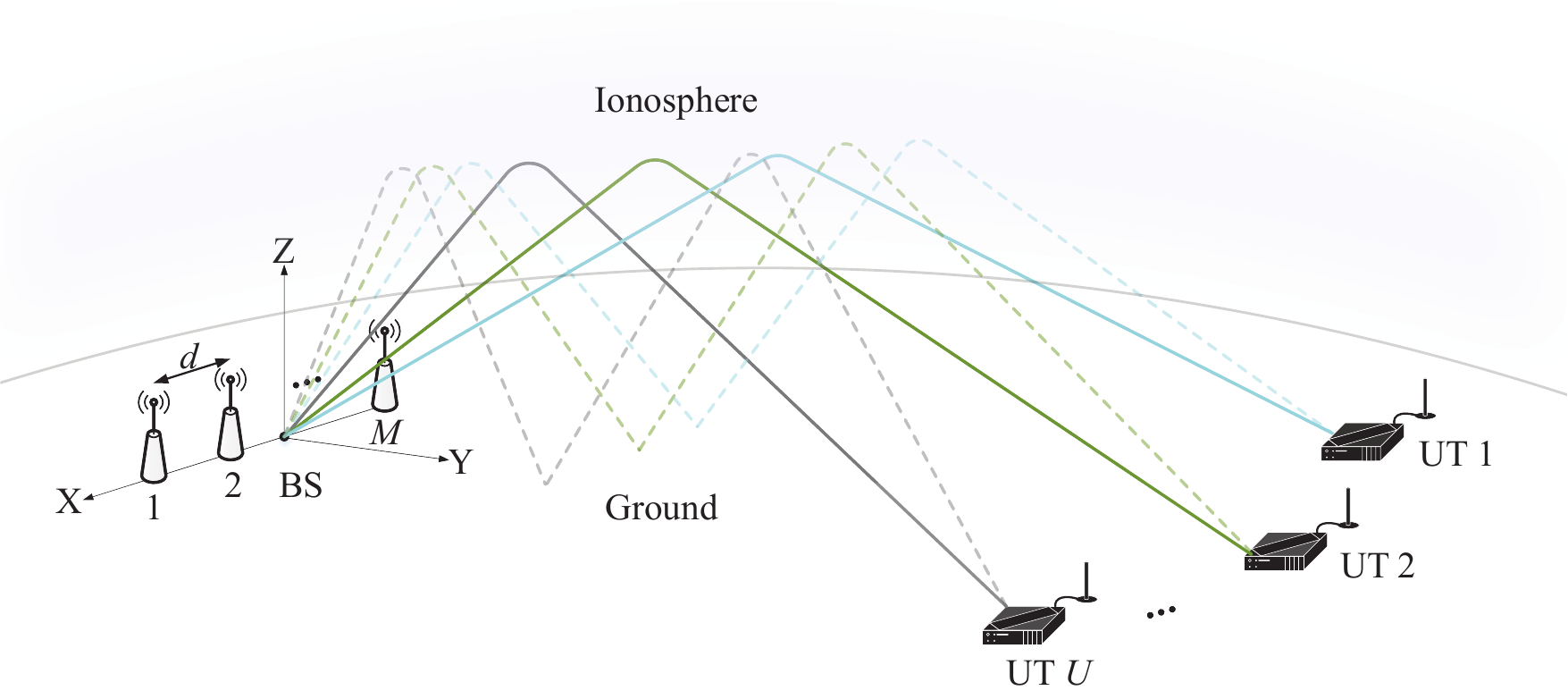}
	\caption{HF skywave massive MIMO system.}
	\label{fig:HF_prop}
\end{figure}

Consider a long-range HF skywave massive MIMO system with orthogonal frequency division multiplexing (OFDM) modulation, as illustrated in Fig.~\ref{fig:HF_prop}. The BS is equipped with $M$-antenna uniform linear array (ULA), serving $U$ single-antenna UTs. Due to the decametric wavelength of the HF waveband, the antenna array aperture of the BS is typically large, resulting in the impracticable of multiple antennas in the elevation direction, different from terrestrial cellular systems. Besides, the ULA configurations have been implemented in the HF skywave over-the-horizon MIMO radar~\cite{8263323}, where hundreds of antennas are deployed. Hence, the implementation of ULA in HF skywave massive MIMO communications is feasible.

In contrast to traditional terrestrial cellular communications, HF skywave communication systems require the carrier frequency $f_{\mathrm{c}}$ to be adjusted accordingly with the complicated ionospheric environment variations~\cite{qin2017link}. Therefore, instead of using the carrier frequency, we use the highest system operating frequency $f_{\mathrm{o}}$ with the wavelength $\lambda_{\mathrm{o}}=c/f_{\mathrm{o}}$ to design antenna arrays, where $c$ is the speed of light. The inter-antenna spacing is set as $d = \lambda_{\mathrm{o}}/2$, which is shorter than half wavelength of carrier, different from that widely used in traditional massive MIMO communications.

\subsection{Modified Beam Based Channel Model}\label{2-B}
For uplink transmission, $U$ complex symbols from different UTs are transmitted over the channel at the same time and received at the BS with $M$-dimensional signal. For each subcarrier of OFDM symbols, the received signal vector can be expressed as 
\begin{equation}\label{sig_mod}
   \mathbf{y} = \mathbf{Hx} + \mathbf{z},
\end{equation}
where indexes of OFDM symbol and subcarrier are omitted for brevity, since the receiver design is individually implemented for each OFDM symbol and subcarrier. In~\eqref{sig_mod}, $\mathbf{y}\in\mathbb{C}^{M}$ is the received signal vector at the BS, $\mathbf{H} = [\mathbf{h}_{1},\cdots,\mathbf{h}_{U}]\in\mathbb{C}^{M\times U}$ is the channel matrix with each column representing the channel response vector for each UT, $\mathbf{x} = [x_{1},\cdots,x_{U}]\in\mathbb{S}^{U}$ is the transmitted signal vector from $U$ UTs and $\mathbb{S}$ denotes the signal constellation, and $\mathbf{z}\sim\mathcal{CN}(\mathbf{0}_{M}, \sigma_{\mathrm{z}}\mathbf{I}_{M})$ is the additive white Gaussian noise.

According to multi path channel model, the channel vector $\mathbf{h}_{u}$ for UT $u$ can be expressed as~\cite{10122719} 
\begin{equation}
   \mathbf{h}_{u} = \sum_{p=1}^{P_{u}}\alpha_{u,p}\mathbf{v}(\Omega_{u,p}),
\end{equation}
where $P_{u}$ is the path number from UT $u$ to the BS, $\alpha_{u,p}$ and $\Omega_{u,p}$ are the complex gain and the directional cosine of the $p$-th path from UT $u$, respectively. The steering vector $\mathbf{v}(\Omega)\in\mathbb{C}^{M}$ pointing towards $\Omega$ is defined as conjugate centrosymmetric structure with the $m$-th element being~\cite{7959626}
\begin{equation}
   [\mathbf{v}(\Omega)]_{m} = \frac{1}{\sqrt{M}}\exp\{-\overline{\jmath}\pi f_{\mathrm{c}}\Delta_{\tau}(2m-M-1)\Omega\},
\end{equation}
where $\Delta_{\tau} = d/c$. 

Owning to the small angle spread of the HF skywave communications~\cite{8904116}, the channel can be represented in the beam domain with remarkable sparsity by performing uniform sampling for directional cosine. In previous works, the beam based channel model is established by sampling the range of directional cosine $[-1,1)$ uniformly, and the sampling interval length is determined by the sampling number~\cite{yu2021hf, HFmimo}. For the scenarios where the inter-antenna spacing is half wavelength of carrier frequency, e.g. traditional terrestrial cellular communications, the beam matrix (constructed by sampled steering vectors) can be expressed as DFT matrix or partial DFT matrix~\cite{9910031}, which enables efficient signal processing by integrating fast Fourier transform (FFT). However, as mentioned in \ref{2-A}, the inter-antenna spacing of HF skywave massive MIMO is usually shorter than half wavelength of carrier, in which case the beam matrix is modelled as chirp z-transform (CZT)-based matrix rather than DFT-based matrix~\cite{HFmimo}, which limits the conventional FFT based efficient implementation of various transmission schemes.

\begin{figure}[ht]
	\centering
	\includegraphics[width=0.48\textwidth]{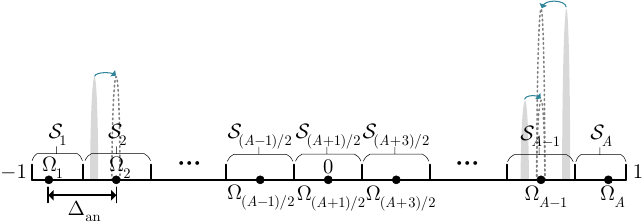}
	\caption{Directional cosine sampling.}
	\label{fig:samp}
\end{figure}

To tackle this problem, we turn to perform uniform sampling for directional cosine with deterministic interval length
\begin{equation}
   \Delta_{\mathrm{an}} = \frac{1}{FMf_{\mathrm{c}}\Delta_{\tau}} = \frac{2}{FM_{\mathrm{eq}}},
\end{equation}
where $F$ is the fine factor, 
and $M_{\mathrm{eq}} \triangleq Mf_{\mathrm{c}}/f_{\mathrm{o}}$ is the equivalent number of BS antennas. For fixed antenna array aperture $d = \lambda_{\mathrm{o}}/2$ and antenna number $M$, the angular resolution of the array is proportional to carrier frequency $f_{\mathrm{c}}$, thus the term $M_{\mathrm{eq}}$ signifies the equivalent number of antennas required to achieve the angular resolution corresponding to the half-wavelength array aperture at any given carrier frequency, and $FM_{\mathrm{eq}}$ reflects the sampling density. 
Consider a symmetric partition of the range of directional cosine $[-1,1)$, we can obtain $A = 2\left\lfloor \Delta_{\mathrm{an}}^{-1}\right\rfloor + 1$ mutually exclusive subsets and the $a$-th subset is defined as follows:
\begin{equation}
   \mathcal{S}_{a}\triangleq
   \left\{\begin{aligned}
      &[-1, \Omega_{a} + \Delta_{\mathrm{an}}/2),&& a = 1\\
      &[\Omega_{a} - \Delta_{\mathrm{an}}/2, \Omega_{a} + \Delta_{\mathrm{an}}/2),&& 1 < a < A\\
      &[\Omega_{a} - \Delta_{\mathrm{an}}/2, 1),&& a = A
       \end{aligned}\right.,
\end{equation}
where $\Omega_{a}\triangleq \left(a - 1 - \left\lfloor \Delta_{\mathrm{an}}^{-1}\right\rfloor\right)\Delta_{\mathrm{an}}$ is the $a$-th sampled directional cosine. Denote set $\mathcal{P}_{u} \triangleq \{\Omega_{u,1}, \cdots,\Omega_{u,P_{u}}\}$ containing directional cosines of all the paths for UT $u$, and map the directional cosine in $\mathcal{S}_{a}$ to $\Omega_{a}$. Fig. \ref{fig:samp} illustrates the schematic diagram of directional cosine sampling, each path is mapped to the nearest sampled grid point. It can be checked that $\Omega_{u,p}$ is mapped to the $a(\Omega_{u,p})$-th sampled directional cosine, where $a(\Omega) \triangleq \lceil\Delta_{\mathrm{an}}^{-1}\Omega\rfloor + \left\lfloor\Delta_{\mathrm{an}}^{-1}\right\rfloor + 1$ is the mapping function.

With the uniform sampling for directional cosine, space domain channel $\mathbf{h}_{u}$ for UT $u$ can be approximated as~\cite{yu2021hf} 
\begin{equation}\label{equ_h_u}
   \mathbf{h}_{u} \approx \sum_{a=1}^{A}\sum_{\Omega_{u,p}\in \mathcal{P}_{u}\cap \mathcal{S}_{a}}\alpha_{u, p}\mathbf{v}_{a} = \mathbf{V}\mathbf{g}_{u},
\end{equation}
where $\mathbf{v}_{a} \triangleq \mathbf{v}(\Omega_{a})\in\mathbb{C}^{M}$ is the sampled steering vector and $\mathbf{V} \triangleq \left[\mathbf{v}_{1}, \cdots,\mathbf{v}_{A}\right]\in\mathbb{C}^{M\times A}$ is termed as the beam matrix, $\mathbf{g}_{u}\in\mathbb{C}^{A}$ is the beam domain channel vector with $[\mathbf{g}_{u}]_{a} = \sum_{\Omega_{u,p}\in \mathcal{P}_{u}\cap \mathcal{S}_{a}}\alpha_{u, p}$. According to the definition, the beam matrix can be expressed as 
\begin{align}\label{V_mtx}
   \mathbf{V} &= \frac{1}{\sqrt{M}}\widetilde{\bm{\Omega}}\mathbf{I}_{M, S}\mathbf{F}_{S}\mathbf{I}_{S, A}\overline{\bm{\Omega}}\nonumber\\
   &= \frac{1}{\sqrt{M}}\mathbf{I}_{M, S}\mathbf{F}_{S}\bm{\Omega}\overline{\bm{\Pi}}_{S}^{S - (A-1)/2}\mathbf{I}_{S, A},
\end{align}
where $S = FM$, $\mathbf{F}_{S} = \left[\mathbf{f}_{S,1},\cdots,\mathbf{f}_{S,S}\right]\in\mathbb{C}^{S\times S}$ represents $S$-dimensional DFT matrix, $\mathbf{I}_{N_{1},N_{2}}$ is composed by the first $N_{1}$ columns ($N_{1}\leq N_{2}$) of $\mathbf{I}_{N_{2}}$ or the first $N_{2}$ rows ($N_{1}>N_{2}$) of $\mathbf{I}_{N_{1}}$,  the phase shift matrices are respectively defined as 
\begin{subequations}
   \begin{align}
      \bm{\Omega} &\triangleq \operatorname{diag}\left\{-\mathbf{I}_{S,2S}\mathbf{f}_{2S,M}^{\ast}\right\}\in\mathbb{C}^{S\times S},\\
      \widetilde{\bm{\Omega}} &\triangleq \operatorname{diag}\left\{\mathbf{I}_{M,S}\mathbf{f}_{S,S-(A-1)/2 + 1}\right\}\in\mathbb{C}^{M\times M},\\
      \overline{\bm{\Omega}} &\triangleq \operatorname{diag}\left\{\mathbf{I}_{A,2S}\mathbf{\Pi}_{2S}^{(A-1)/2}\mathbf{f}_{2S,M}^{\ast}\right\}\in\mathbb{C}^{A\times A}.
   \end{align}
\end{subequations}

Each sampled steering vector in the channel representation \eqref{equ_h_u} is associated with a physical beam in the space domain, and all UTs utilize an identical set of sampled steering vectors. Moreover, according to \eqref{V_mtx}, the beam matrix can be viewed as partial of the permutation of a phase-shifted DFT matrix, which is superior to the CZT-based beam matrix since the proposed beam matrix has a straightforward structure related to DFT. Then~\eqref{equ_h_u} is termed as the modified beam based channel model (BBCM), and can be written in a compact form 
\begin{equation}\label{H}
   \mathbf{H} = \mathbf{VG}. 
\end{equation}
where $\mathbf{G}\triangleq [\mathbf{g}_{1},\cdots,\mathbf{g}_{U}]\in\mathbb{C}^{A\times U}$ is referred to as the beam domain channel matrix.




On account of the small angle spread and the limited propagation path number of HF skywave channels, the beam domain channel vector shows significant sparsity, i.e., the number of non-zero elements of $\mathbf{g}_{u},\ u\in\mathcal{Z}_{U}^{+}$, is relatively small~\cite{yu2021hf}. The non-zero beam set for UT $u$ is given as $\mathcal{A}_{u}\triangleq \left\{a\in\mathcal{Z}_{A}^{+} \big| [\mathbf{g}_{u}]_{a}\neq 0\right\}$ with $A_{u} = |\mathcal{A}_{u}|$, which can also be obtained by $\mathcal{A}_{u}= \left\{a\in\mathcal{Z}_{A}^{+} \big| [\bm{\omega}_{u}]_{a}\neq 0\right\}$, where the beam domain channel coupling vector is defined as 
\begin{equation}
   \bm{\omega}_{u} \triangleq \mathbb{E}\left\{\mathbf{g}_{u}\odot\mathbf{g}_{u}^{\ast}\right\}\in\mathbb{R}^{A}.
\end{equation}
The beam domain channel coupling vector, $\bm{\omega}_{u}$, is typically referred to as the beam domain statistical channel state information (CSI) and varies slowly with time~\cite{yu2021hf}, thus it is efficient to leverage $\bm{\omega}_{u}$ to obtain $\mathcal{A}_{u}$. Moreover, in this paper, we assume that both the beam domain instantaneous CSI $\mathbf{G}$ and the beam domain statistical CSI $\bm{\omega}_{u},\ u\in\mathcal{Z}_{U}^{+}$, are available at the BS.

\section{Beam Structured Turbo Receiver}\label{BSTR}
In this section, we first provide an overview of the turbo receiver. The BSTR is further proposed based on the modified BBCM, where the beam structured signal detection using the a priori information is derived, involving a group-wise low-dimensional beam domain detector design. We show that under certain conditions, the BSTR is asymptotically optimal in the sense of minimizing MSE.
\subsection{An Overview of the Turbo Receiver}\label{TR}
\begin{figure}[ht]
	\centering
	\includegraphics[width=0.48\textwidth]{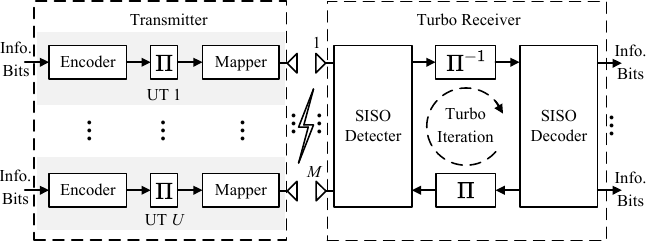}
	\caption{Block diagram of a massive MIMO system with BICM and turbo receiver, where $\Pi$ and $\Pi^{-1}$ represent the interleaver and the deinterleaver, respectively.}
	\label{fig:bicm}
\end{figure}

Typically, turbo receivers are implemented within the bit-interleaved coded modulation (BICM) systems~\cite{10272020}. As depicted in Fig.~\ref{fig:bicm}, the original information bitstream undergoes sequential processing through an encoder and an interleaver at each UT. Subsequently, the interleaved bitstream is partitioned into vectors of length $N$ and mapped to constellation points in $\mathbb{S} = \{s_{1},\cdots,s_{K}\}$ using Gray mapping, where $s_{k}\in\mathbb{C},\ k\in\mathcal{Z}_{K}^{+}$, is the constellation point corresponding to the bit sequence $\mathbf{b}_{k} = [b_{k,1},\cdots,b_{k,N}]^{\mathrm{T}}\in\mathbb{B}^{N}$, and $K = 2^{N}$ is the modulation order, thereby generating the transmitted symbols. The bit sequence mapped to the transmitted symbol $x_{u},\ u\in\mathcal{Z}_{U}^{+}$, is employed as $\mathbf{q}_{u} = [q_{u,1},\cdots,q_{u,N}]^{\mathrm{T}}\in\mathbb{B}^{N}$, where $x_{u}\in\mathbb{S}$ and $\mathbf{q}_{u}\in\{\mathbf{b}_{1},\cdots,\mathbf{b}_{K}\}$. It is natural to assume that the transmitted symbols of all UTs are uniformly and independently drawn from the signal constellation $\mathbb{S}$ at random, and the transmitted power of $U$ symbols are normalized to unit, i.e., $\mathbb{E}\left\{\mathbf{xx}^{\mathrm{H}}\right\} = \mathbf{I}_{U}$. At the BS, the received signal is expressed as~\eqref{sig_mod}, and a turbo receiver is adopted, enabling the iterative exchange of soft information between the SISO detector and the SISO decoder during the turbo iteration, while the decoded information bits are obtained at the final iteration. Specifically, the iteration process of the SISO detector consists of the following steps: 

\subsubsection{Signal Detection using the a Priori Information}
In the turbo iteration process, the a priori mean and variance of the transmitted signal $\mathbf{x}$ are respectively denoted as 
\begin{subequations}\label{pri_inf}
   \begin{align}
      \bm{\mu} &\triangleq \mathbb{E}\{\mathbf{x}\} = [\mu_{1},\cdots,\mu_{U}]^{\mathrm{T}}\in\mathbb{C}^{U},\\
      \bm{\Sigma} &\triangleq \operatorname{Cov}\{\mathbf{x},\mathbf{x}\} = \operatorname{diag}\{\sigma_{1},\cdots,\sigma_{U}\}\in\mathbb{R}^{U\times U}.
   \end{align}
\end{subequations}
The values of $\bm{\mu}$ and $\bm{\Sigma}$ are respectively initialized to $\mathbf{0}_{U}$ and $\mathbf{I}_{U}$, and are updated through the iteration process. With the a priori mean and variance, the signal detection for UT $u$ aims to make an approximation of $p(\mathbf{y}|x_{u} = s_{k}),\ k\in\mathcal{Z}_{K}^{+}$, for the extrinsic information calculation, where $p(\cdot)$ denotes the probability density function of continuous random variables. 
This extrinsic information, after deinterleaving, decoding, and interleaving, is subsequently employed to update the a prior information for the next iteration of detection.

\subsubsection{The Extrinsic Information Calculation}
Based on the received signal $\mathbf{y}$, the LLR of bit $q_{u,i},\ u\in\mathcal{Z}_{U}^{+}, i\in\mathcal{Z}_{N}^{+}$, is given as 
\begin{align}\label{Lq}
   L(q_{u,i}) =& \ln\frac{P(q_{u,i} = 1| \mathbf{y})}{P(q_{u,i} = 0| \mathbf{y})} = \ln\frac{\sum_{s_{k}\in\mathbb{S}_{i}^{1}}P(x_{u} = s_{k}| \mathbf{y})}{\sum_{s_{k}\in\mathbb{S}_{i}^{0}}P(x_{u} = s_{k}| \mathbf{y})}\nonumber\\
   =& \underbrace{\ln\frac{\sum_{s_{k}\in\mathbb{S}_{i}^{1}}p(\mathbf{y} | x_{u} = s_{k}) \prod_{j \neq i}^{N}P(q_{u,j} = b_{k,j})}{\sum_{s_{k}\in\mathbb{S}_{i}^{0}}p(\mathbf{y} | x_{u} = s_{k}) \prod_{j \neq i}^{N}P(q_{u,j} = b_{k,j})}}_{\triangleq L_{\mathrm{e}}(q_{u,i})}\nonumber\\
   &+ \ln\frac{P(q_{u,i} = 1)}{P(q_{u,i} = 0)},
\end{align}
where $P(\cdot)$ represents the probability distribution of discrete random variables, $\mathbb{S}_{i}^{b}\triangleq \{s_{k}\in\mathbb{S} | b_{k,i} = b\},\ b\in\mathbb{B},i\in\mathcal{Z}_{N}^{+}$, is the symbol set with the $i$-th mapping bit $b_{k,j}$ of $s_{k}$ being $b$. Furthermore, the probability $P(q_{u,i} = b),\ b\in\mathbb{B}$, is calculated by the a priori LLR $L_{\mathrm{a}}^{\prime}(q_{u,i})$, and is expressed as $P(q_{u,j} = b) = \frac{1}{2}\left(1+(2b - 1)\tanh\big(L_{\mathrm{a}}^{\prime}(q_{u,j})/2\big)\right)$.
Consequently, the extrinsic LLR $L_{\mathrm{e}}(q_{u,i})$ in~\eqref{Lq} is utilized for SISO decoding. 

\subsubsection{The a Priori Information Update}
After deinterleaving, $L_{\mathrm{e}}(q_{u,i})$ is mapped to $L_{\mathrm{a}}(c_{u,i})$, where $c_{u,i} = \Pi^{-1}(q_{u,i})$ and $\Pi(\cdot)$ is the interleaver. Let $L_{\mathrm{a}}(c_{u,i})$ denote the a prior soft information for SISO decoder's input, which, after the decoding process, yields $L_{\mathrm{e}}^{\prime}(c_{u,i})$. Subsequently, $L_{\mathrm{a}}^{\prime}(q_{u,i})$ is obtained after interleaving, which is then employed as the a prior soft information for the detection input. Based on $L_{\mathrm{a}}^{\prime}(q_{u,i})$, the updated symbol probability is $P(x_{u} = s_{k}) =\prod_{j = 1}^{N}\frac{1}{2}\left(1+(2b_{k,j} - 1)\tanh\big(L_{\mathrm{a}}^{\prime}(q_{u,j})/2\big)\right)$.
The mean and variance of the transmitted symbol are respectively updated as 
$\mu_{u} = \sum_{k=1}^{K}s_{k}P(x_{u} = s_{k})$ and $\sigma_{u} = \sum_{k=1}^{K}|s_{k}|^{2}P(x_{u} = s_{k}) - |\mu_{u}|^{2}$,
which are used as the input in~\eqref{pri_inf} for the next iteration. Moreover, $L_{\mathrm{a}}^{\prime}(q_{u,i})$ is also adopted for the calculation of $L_{\mathrm{e}}(q_{u,i})$ at the next iteration.

\subsection{Beam Structured Turbo Receiver}
\label{BSD_pri}

We first focus on the signal detection using the a priori information. To develop a low-complexity signal detection approach, we consider linear detection combined with the modified BBCM. 
Before proceeding, we categorize $U$ UTs into $L$ disjoint groups $\mathcal{N}_{1},\cdots,\mathcal{N}_{L}$ with $N_{l} = |\mathcal{N}_{l}|,\ l\in\mathcal{Z}_{L}^{+}$, and define the UT selection matrix $\mathbf{N}_{l}\triangleq\mathbf{S}(U, \mathcal{N}_{l})\in\mathbb{R}^{U\times N_{l}},\ l\in\mathcal{Z}_{L}^{+}$. 
Then the detection of $\mathbf{x}$ can be decoupled into $L$ groups, where the detection for the $l$-th group's transmitted signal $\mathbf{x}_{l} \triangleq \mathbf{N}_{l}^{\mathrm{T}}\mathbf{x}\in\mathbb{C}^{N_{l}}$ can be given as 
\begin{equation}\label{hat_xl}
   \widehat{\mathbf{x}}_{l} = \mathbf{R}_{l}^{\mathrm{H}}\mathbf{y} + \mathbf{n}_{l}.
\end{equation} 
The MSE for the detection of $\mathbf{x}_{l}$ is given by $\epsilon_{l}(\mathbf{R}_{l}, \mathbf{n}_{l}) \triangleq \mathbb{E}\left\{\left\|\mathbf{x}_{l}-\left(\mathbf{R}_{l}^{\mathrm{H}}\mathbf{y} + \mathbf{n}_{l}\right)\right\|_{2}^{2}\right\}$. In terms of minimizing MSE, the optimal $\mathbf{R}_{l}$ and $\mathbf{n}_{l}$ can be obtained by
\begin{equation}\label{rl_dl_op}
   \mathbf{R}_{l}^{\mathrm{op}}, \mathbf{n}_{l}^{\mathrm{op}} = \underset{\mathbf{R}_{l}\in\mathbb{C}^{M\times N_{l}},\mathbf{n}_{l}\in\mathbb{C}^{N_{l}}}{\operatorname{argmin}} \epsilon_{l}(\mathbf{R}_{l}, \mathbf{n}_{l}),
\end{equation}
Solving~\eqref{rl_dl_op} yields the MMSE detector with the a priori information~\cite{984761}
\begin{subequations}\label{R_d}
   \begin{align}
      \mathbf{R}_{l}^{\mathrm{op}} 
      &= \mathbf{H}\left(\bm{\Sigma}\mathbf{H}^{\mathrm{H}}\mathbf{H} + \sigma_{\mathrm{z}}\mathbf{I}\right)^{-1}\bm{\Sigma}\mathbf{N}_{l},\\
      \mathbf{n}_{l}^{\mathrm{op}} 
      &= \mathbf{N}_{l}^{\mathrm{T}}\bm{\mu} - \left(\mathbf{R}^{\mathrm{op}}_{l}\right)^{\mathrm{H}}\mathbf{H}\bm{\mu}.
   \end{align}
\end{subequations}
For massive MIMO systems, the computational complexity of~\eqref{R_d} will become prohibitively high as the antenna number of the BS and the UT number rise, due to the large-dimensional matrix inversion and multiplication. To this end, we turn to exploring intrinsic structures of the optimal solution of~\eqref{rl_dl_op}. 


Define the beam set and the beam selection matrix for group $l$ as $\mathcal{B}_{l}\triangleq \cup_{u\in\mathcal{N}_{l}}\mathcal{A}_{u}$ and $\mathbf{B}_{l} \triangleq \mathbf{S}(A,\mathcal{B}_{l})\in\mathbb{R}^{A\times B_{l}}$ with $B_{l} = |\mathcal{B}_{l}|$, respectively, and assume that the UT number is limited and the directional cosines of the paths from different UTs are distinct~\cite{yu2021hf}, then we have the following theorem, proved in Appendix~\ref{apd2}.

\begin{theorem}\label{theo1}
   When $\forall l\neq l^{\prime},\ \mathcal{B}_{l}\cap\mathcal{B}_{l^{\prime}} = \varnothing$, as $M\rightarrow\infty$, the optimal solution of~\eqref{rl_dl_op} can be obtained as 
   \begin{subequations}\label{ru_du1}
      \begin{align}
         \mathbf{R}_{l}^{\mathrm{op}} &= \mathbf{V}\mathbf{B}_{l}\mathbf{W}_{l}^{\mathrm{op}},\\
         \mathbf{n}_{l}^{\mathrm{op}} &= \bm{\mu}_{l} - \left(\mathbf{W}_{l}^{\mathrm{op}}\right)^{\mathrm{H}}\mathbf{B}_{l}^{\mathrm{T}}\mathbf{G}\bm{\mu},
      \end{align}
   \end{subequations}
   where $\bm{\mu}_{l}\triangleq \mathbf{N}_{l}^{\mathrm{T}}\bm{\mu}\in\mathbb{C}^{B_{l}}$, $\bm{\Sigma}_{l}\triangleq\mathbf{N}_{l}^{\mathrm{T}}\bm{\Sigma}\mathbf{N}_{l}\in\mathbb{R}^{B_{l}\times B_{l}}$, and
   \begin{align}\label{wu_til}
      \mathbf{W}_{l}^{\mathrm{op}} =& \underset{\mathbf{W}_{l}\in\mathbb{C}^{B_{l}\times N_{l}}}{\operatorname{argmin}}\operatorname{tr}\Big\{\mathbf{W}_{l}^{\mathrm{H}}\left(\mathbf{B}_{l}^{\mathrm{T}}\mathbf{G}\bm{\Sigma}\mathbf{G}^{\mathrm{H}}\mathbf{B}_{l} + \sigma_{\mathrm{z}}\mathbf{I}\right)\mathbf{W}_{l}\nonumber\\
      &- \bm{\Sigma}_{l}\mathbf{N}_{l}^{\mathrm{T}}\mathbf{G}^{\mathrm{H}}\mathbf{B}_{l}\mathbf{W}_{l} - \mathbf{W}_{l}^{\mathrm{H}}\mathbf{B}_{l}^{\mathrm{T}}\mathbf{G}\mathbf{N}_{l}\bm{\Sigma}_{l} + \bm{\Sigma}_{l}\Big\}.
   \end{align}
\end{theorem}

According to Theorem~\ref{theo1}, to minimize MSE, the asymptotically optimal space domain detector is beam structured. Although the number of BS antennas is finite in practical systems, Theorem~\ref{theo1} can still guide the detector design when $M$ is sufficiently large. Specifically, for group $l$, we restrict the space domain detector $\mathbf{R}_{l}$ as beam structure
\begin{equation}\label{r_cons1}
   \mathbf{R}_{l} = \mathbf{V}\mathbf{B}_{l}\mathbf{W}_{l},
\end{equation}
where $\mathbf{W}_{l}\in\mathbb{C}^{B_{l}\times N_{l}}$ is the beam domain detector for group $l$, then~\eqref{hat_xl} can be expressed as
\begin{equation}\label{xl_hat_bs}
   \widehat{\mathbf{x}}_{l} = \mathbf{W}_{l}^{\mathrm{H}}\mathbf{B}_{l}^{\mathrm{T}}\mathbf{V}^{\mathrm{H}}\mathbf{y} + \mathbf{n}_{l} = \mathbf{W}_{l}^{\mathrm{H}}\mathbf{y}_{l} + \mathbf{n}_{l},
\end{equation} 
where $\mathbf{y}_{l}\triangleq \mathbf{B}_{l}^{\mathrm{T}}\mathbf{V}^{\mathrm{H}}\mathbf{y}\in\mathbb{C}^{B_{l}}$.
Then the MSE for the detection of group $l$ is $\overline{\epsilon}_{l}(\mathbf{W}_{l}, \mathbf{n}_{l}) =\mathbb{E}\left\{\left\|\mathbf{x}_{l}-\left(\mathbf{W}_{l}^{\mathrm{H}}\mathbf{y}_{l} + \mathbf{n}_{l}\right)\right\|_{2}^{2}\right\}$.
By minimizing $\overline{\epsilon}_{l}(\mathbf{W}_{l}, \mathbf{n}_{l})$, we have~\cite{984761} 
\begin{subequations}\label{Wl-dl}
   \begin{align}
      \mathbf{W}_{l}&= \operatorname{Cov}\{\mathbf{y}_{l},\mathbf{y}_{l}\}^{-1}\operatorname{Cov}\{\mathbf{y}_{l},\mathbf{x}_{l}\} \nonumber\\
      &= \left(\mathbf{D}_{l}\bm{\Sigma}\mathbf{D}_{l}^{\mathrm{H}} + \sigma_{\mathrm{z}}\mathbf{B}_{l}^{\mathrm{T}}\mathbf{Q}\mathbf{B}_{l}\right)^{-1}\widetilde{\mathbf{D}}_{l}\bm{\Sigma}_{l},\label{Wl}\\
      \mathbf{n}_{l}&= \mathbb{E}\{\mathbf{x}_{l}\} - \operatorname{Cov}\{\mathbf{x}_{l},\mathbf{y}_{l}\}\operatorname{Cov}\{\mathbf{y}_{l},\mathbf{y}_{l}\}^{-1}\mathbb{E}\{\mathbf{y}_{l}\} \nonumber\\
      &= \bm{\mu}_{l} - \mathbf{W}_{l}^{\mathrm{H}}\mathbf{D}_{l}\bm{\mu},
   \end{align}
\end{subequations}
where $\mathbf{D}_{l} \triangleq \mathbf{B}_{l}^{\mathrm{T}}\mathbf{D}\in\mathbb{C}^{B_{l}\times U}$, $\widetilde{\mathbf{D}}_{l}\triangleq \mathbf{D}_{l}\mathbf{N}_{l}\in\mathbb{C}^{B_{l}\times N_{l}}$, and 
\begin{subequations}\label{D_Q}
   \begin{align}
      \mathbf{D}&\triangleq\mathbf{QG}\in\mathbb{C}^{A\times U},\\
      \mathbf{Q}&\triangleq\mathbf{V}^{\mathrm{H}}\mathbf{V}\in\mathbb{C}^{A\times A}.
   \end{align}
\end{subequations}
Consequently, the detection for group $l$ is given as
\begin{equation}\label{xhat_l}
   \widehat{\mathbf{x}}_{l} = \mathbf{W}_{l}^{\mathrm{H}}\mathbf{y}_{l} + \mathbf{n}_{l} = \mathbf{W}_{l}^{\mathrm{H}}\widetilde{\mathbf{y}}_{l} + \bm{\mu}_{l},
\end{equation}
where $\widetilde{\mathbf{y}}_{l}\triangleq \mathbf{B}_{l}^{\mathrm{T}}\widetilde{\mathbf{y}}\in\mathbb{C}^{B_{l}}$ is a subvector of 
\begin{equation}\label{y_til}
   \widetilde{\mathbf{y}}= \mathbf{V}^{\mathrm{H}}(\mathbf{y} - \overline{\mathbf{y}}),
\end{equation}
in which $\overline{\mathbf{y}}\triangleq \mathbf{VG}\bm{\mu}\in\mathbb{C}^{M}$.

In practice, to achieve satisfactory performance with low complexity, appropriate UT scheduling and grouping strategies can be employed to minimize the cardinality of the beam set of each UT group. 
Given the pronounced sparsity of the beam domain channel and the grouping scheme, the computations of~\eqref{Wl-dl} and~\eqref{xhat_l}, the so-called beam domain detector design and beam domain signal detection, only involve low-dimensional operations, resulting in relatively low complexity. 
Moreover, the beam structured signal detector in~\cite{BSD} can be viewed as a special case of the proposed grouping-based iterative approach, namely, when each group contains only one UT and there is no iterative update on the a prior information.

After the beam structured signal detection using the a priori information, the extrinsic information calculation can be effectively implemented. Assuming $\mathbf{x}\sim\mathcal{CN}(\bm{\mu}, \bm{\Sigma})$, for $u\in\mathcal{N}_{l},\ l\in\mathcal{Z}_{L}^{+}$, we have~\cite{5741768} 
\begin{equation}
   p(\mathbf{y}_{l}|x_{u}) = \frac{p(x_{u}|\mathbf{y}_{l})p(\mathbf{y}_{l})}{p(x_{u})}\propto\exp\left\{-\frac{|x_{u} - \mu_{u}^{\mathrm{e}}|^{2}}{\sigma_{u}^{\mathrm{e}}}\right\},
\end{equation}
where the extrinsic mean and the extrinsic variance of $x_{u}$ are expressed as 
\begin{subequations}
   \begin{align}
      \mu_{u}^{\mathrm{e}} &= (\mu_{u}^{\mathrm{p}}/\sigma_{u}^{\mathrm{p}} - \mu_{u}/\sigma_{u})\sigma_{u}^{\mathrm{e}},\\
      \sigma_{u}^{\mathrm{e}} &= (1/\sigma_{u}^{\mathrm{p}} - 1/\sigma_{u})^{-1},
   \end{align}
\end{subequations}
respectively, and 
\begin{subequations}
   \begin{align}
      \mu_{u}^{\mathrm{p}} &= \mathbb{E}\{x_{u}|\mathbf{y}_{l}\} = \mathbf{e}_{u}^{\mathrm{T}}\mathbf{N}_{l}\widehat{\mathbf{x}}_{l},\\
      \sigma_{u}^{\mathrm{p}} &= \operatorname{Cov}\{x_{u}, x_{u}|\mathbf{y}_{l}\} = \left(1 - \mathbf{e}_{u}^{\mathrm{T}}\mathbf{N}_{l}\mathbf{W}_{l}^{\mathrm{H}}\mathbf{D}_{l}\mathbf{e}_{u}\right)\sigma_{u},
   \end{align}
\end{subequations}
are denoted as the a posteriori mean and the a posteriori variance of $x_{u}$, respectively. By approximating $p(\mathbf{y}|x_{u})$ as $p(\mathbf{y}_{l}|x_{u})$, the extrinsic LLR $L_{\mathrm{e}}(q_{u,i})$ can be computed by 
\begin{align}\label{Leq}
   &L_{\mathrm{e}}(q_{u,i}) = \nonumber\\
   &\ln\!\frac{\sum_{s_{k}\in\mathbb{S}_{i}^{1}}\!\exp\!\left\{\!-|\mu_{u}^{\mathrm{e}}\! - \!s_{k}|^{2}/\sigma_{u}^{\mathrm{e}} \!+\! \sum_{j\neq i}^{N} (b_{k,j}\! - \!1/2)L_{\mathrm{a}}^{\prime}(q_{u,j})\!\right\}}{\sum_{s_{k}\in\mathbb{S}_{i}^{0}}\!\exp\!\left\{\!-|\mu_{u}^{\mathrm{e}}\! - \!s_{k}|^{2}/\sigma_{u}^{\mathrm{e}} \!+\! \sum_{j\neq i}^{N} (b_{k,j}\! -\! 1/2)L_{\mathrm{a}}^{\prime}(q_{u,j})\!\right\}}.
\end{align}

\begin{figure*}[ht]
	\centering
	\includegraphics[width=1\textwidth]{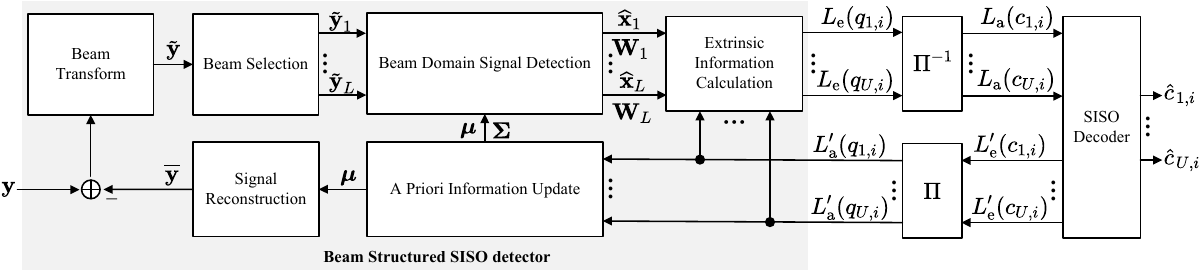}
	\caption{Block diagram of the BSTR.}
	\label{fig:turbo_idd_new}
\end{figure*}

In addition to the beam structured signal detection and the extrinsic information calculation, the a priori information update stays consistent with that described in~\ref{TR}, in combination with the SISO decoder, thereby forming the beam structured turbo receiver (BSTR). 
Fig.~\ref{fig:turbo_idd_new} concludes the proposed BSTR by a block diagram. The beam transform is first performed on the demeaned signal $\mathbf{y}-\overline{\mathbf{y}}$, resulting in $\widetilde{\mathbf{y}}$, where $\overline{\mathbf{y}}$ is set as $\mathbf{0}_{M}$ for the first iteration and obtained through the signal reconstruction with the updated a priori mean $\bm{\mu}$ in the subsequent iterations. Next, the beam selection operation yields the beam domain signals $\widetilde{\mathbf{y}}_{l},\ l\in\mathcal{Z}_{L}^{+}$. Then, the beam domain signal detection, the extrinsic information calculation, and the a priori information update are performed to complete the beam structured SISO detector. The soft information is exchanged between the SISO decoder and the beam structured SISO detector with interleaving and deinterleaving at each iteration while the decoded information bits $\widehat{c}_{u,i},\ u\in\mathcal{Z}_{U}^{+}, i\in\mathcal{Z}_{N}^{+}$, are acquired by the SISO decoder at the final iteration. 
Moreover, efficient implementation of the BSTR will be discussed in Section~\ref{WBSTR}.

\section{BSTR with Windowing}\label{WBSTR}
In this section, we generalize the BSTR to the windowed version for further complexity reduction. We design an energy-focusing window for windowed BSTR. Next, efficient implementation of windowed BSTR is investigated and the computational complexity is analyzed.
\subsection[short]{Windowed BSTR}
From~\eqref{xl_hat_bs}, the beam domain signal detection of group $l$ is fundamentally congruent with the signal detection based on the reduced-dimensional signal model:
\begin{equation}
   \mathbf{y}_{l} = \mathbf{D}_{l}\mathbf{x} + \mathbf{z}_{l} = \widetilde{\mathbf{D}}_{l}\mathbf{x}_{l} + \widetilde{\mathbf{D}}_{(l)}\mathbf{x}_{(l)} + \mathbf{z}_{l},
\end{equation}
where $\mathbf{z}_{l}\triangleq \mathbf{B}_{l}^{\mathrm{T}}\mathbf{V}^{\mathrm{H}}\mathbf{z}\in\mathbb{C}^{B_{l}}$, $\widetilde{\mathbf{D}}_{(l)}\triangleq \mathbf{D}_{l}\mathbf{N}_{(l)}\in\mathbb{C}^{B_{l}\times(U-N_{l})}$ and $\mathbf{x}_{(l)}\triangleq \mathbf{N}_{(l)}^{\mathrm{T}}\mathbf{x}\in\mathbb{S}^{U-N_{l}}$ with $\mathbf{N}_{(l)}\triangleq \mathbf{S}(U, \mathcal{Z}_{U}^{+}\setminus\mathcal{N}_{l})\in\mathbb{R}^{U\times (U-N_{l})}$. When directional cosines of different UT groups are all different, the interference term $\widetilde{\mathbf{D}}_{(l)}\mathbf{x}_{(l)}$ vanishes due to the asymptotic orthogonality of the sampled steering vectors as $M\rightarrow\infty$~\cite{BDRP}. However, in reality, the finite antenna number at the BS leads to the non-orthogonality of the sampled steering vectors, which brings remarkable energy diffusion of $\mathbf{D}$ thus amplifies the effect of inter-group interference $\widetilde{\mathbf{D}}_{(l)}\mathbf{x}_{(l)}$. To alleviate this, we incorporate a window function $\bm{\Lambda}\triangleq \operatorname{diag}\{\bm{\eta}\}\in\mathbb{C}^{M\times M}$ with $\bm{\eta} = [\eta_{1}, \cdots, \eta_{M}]^{\mathrm{T}}\in\mathbb{C}^{M}$ to enhance the energy concentration within $\mathbf{D}$.

For the detection of $\mathbf{x}_{l},\ l\in\mathcal{Z}_{L}^{+}$, instead of~\eqref{hat_xl}, we apply the detector to the windowed received signal $\bm{\Lambda}\mathbf{y}$ by 
\begin{equation}
   \widehat{\mathbf{x}}_{l}^{\mathrm{win}} = \widehat{\mathbf{R}}_{l}^{\mathrm{H}}\bm{\Lambda}\mathbf{y} + \widehat{\mathbf{n}}_{l}.
\end{equation}
Similar to \eqref{r_cons1}, we further constrain the detector of group $l$ to have beam structure $\widehat{\mathbf{R}}_{l} = \mathbf{V}\mathbf{B}_{l}\widehat{\mathbf{W}}_{l}$. By minimizing the MSE and in a manner analogous to the derivation process in~\ref{BSD_pri}, the detection of $\mathbf{x}_{l}$ can be obtained as 
\begin{equation}\label{xhat_l_win}
   \widehat{\mathbf{x}}_{l}^{\mathrm{win}} = \widehat{\mathbf{W}}_{l}^{\mathrm{H}}\widehat{\mathbf{y}}_{l} + \bm{\mu}_{l},
\end{equation}
where $\widehat{\mathbf{y}}_{l}\triangleq\mathbf{B}_{l}^{\mathrm{T}}\widehat{\mathbf{y}}\in\mathbb{C}^{B_{l}}$ is a subvector of 
\begin{equation}\label{y_hat0}
   \widehat{\mathbf{y}}= \mathbf{V}^{\mathrm{H}}\bm{\Lambda}(\mathbf{y} - \overline{\mathbf{y}}),
\end{equation}
and the windowed beam domain detector
\begin{equation}\label{Wl_hat}
   \widehat{\mathbf{W}}_{l} = \left(\widehat{\mathbf{D}}_{l}\bm{\Sigma}\widehat{\mathbf{D}}_{l}^{\mathrm{H}} + \sigma_{\mathrm{z}}\mathbf{U}_{l}\right)^{-1}\overline{\mathbf{D}}_{l}\bm{\Sigma}_{l},
\end{equation}
where $\widehat{\mathbf{D}}_{l}\triangleq\mathbf{B}_{l}^{\mathrm{T}}\widehat{\mathbf{D}}\in\mathbb{C}^{B_{l}\times U}$, $\overline{\mathbf{D}}_{l}\triangleq\widehat{\mathbf{D}}_{l}\mathbf{N}_{l}\in\mathbb{C}^{B_{l}\times N_{l}}$, $\mathbf{U}_{l}\triangleq \mathbf{B}_{l}^{\mathrm{T}}\mathbf{U}\mathbf{B}_{l}\in\mathbb{R}^{B_{l}\times B_{l}}$, and 
\begin{subequations}\label{D_hat}
   \begin{align}
      \widehat{\mathbf{D}}&\triangleq\widehat{\mathbf{Q}}\mathbf{G}\in\mathbb{C}^{A\times U},\\
      \widehat{\mathbf{Q}}&\triangleq \mathbf{V}^{\mathrm{H}}\bm{\Lambda}\mathbf{V}\in\mathbb{C}^{A\times A},\\
      \mathbf{U}&\triangleq \mathbf{V}^{\mathrm{H}}\bm{\Lambda}\bm{\Lambda}^{\mathrm{H}}\mathbf{V}\in\mathbb{C}^{A\times A}.
   \end{align}
\end{subequations}

The BSTR with~\eqref{xhat_l} replaced by~\eqref{xhat_l_win} is referred to as the windowed BSTR. Compared with $\mathbf{D}$, the energy of $\widehat{\mathbf{D}}$ can be more concentrated by properly selecting the window function $\bm{\Lambda}$, thus facilitates interference suppression and complexity reduction of the windowed BSTR. Moreover, the windowed BSTR will be reduced to the BSTR in Section~\ref{BSTR} when $\bm{\Lambda} = \mathbf{I}_{M}$, i.e., the rectangular window is applied, in which case $\widehat{\mathbf{y}} = \widetilde{\mathbf{y}}$ and $\widehat{\mathbf{Q}} = \mathbf{U} = \mathbf{Q}$.
\subsection{Energy-Focusing Window Design}
To design the window function which maximizes the energy concentration of $\widehat{\mathbf{D}}$, without loss of generality, we consider the one-path channel $\mathbf{h}(\Omega) = \alpha\mathbf{v}(\Omega)\in\mathbb{C}^{M}$ with complex gain $\alpha$ and directional cosine $\Omega$ due to the linearity of multipath components. Since $\widehat{\mathbf{D}}$ can be expressed as $\widehat{\mathbf{D}} = \mathbf{V}^{\mathrm{H}}\bm{\Lambda}\mathbf{H}$, we further define the beam transform of $\mathbf{h}(\Omega)$ after windowing as 
\begin{equation}\label{d}
   \widehat{\mathbf{d}}(\Omega,\bm{\eta}) \triangleq \mathbf{V}^{\mathrm{H}}\bm{\Lambda}\mathbf{h}(\Omega) = \alpha\mathbf{V}^{\mathrm{H}}\operatorname{diag}\{\bm{\eta}\}\mathbf{v}(\Omega).
\end{equation}
Our objective is to determine a so-called energy-focusing window that enables the energy of $\widehat{\mathbf{d}}(\Omega,\bm{\eta})$ to be focused around the sampled point $a(\Omega)$. Let the index set for which the desired energy concentration is maximized be denoted as $\mathcal{C}(\Omega)\triangleq \{a(\Omega) - c, \cdots, a(\Omega), \cdots, a(\Omega) + c\}$ with a preset integer $c$. For $\widehat{\mathbf{d}}(\Omega,\bm{\eta})$, the total energy and the energy inside region $\mathcal{C}(\Omega)$ can be defined as 
\begin{subequations}
   \begin{align}
      P_{\mathrm{A}}(\Omega,\bm{\eta}) &\triangleq \big\|\widehat{\mathbf{d}}(\Omega,\bm{\eta})\big\|_{2}^{2},\\
      P_{\mathrm{C}}(\Omega,\bm{\eta}) &\triangleq \big\|\mathbf{S}^{\mathrm{T}}(A, \mathcal{C}(\Omega))\widehat{\mathbf{d}}(\Omega,\bm{\eta})\big\|_{2}^{2},
   \end{align}
\end{subequations} 
respectively. 
In order to design a unified window function for different channels, we assume that the directional cosines of UTs within a specific sector are uniformly distributed in $[-\Omega^{\prime}, \Omega^{\prime}]$ with $0<\Omega^{\prime}\leq 1$ and define the average energy ratio~\cite{8834790}
\begin{equation}
   \lambda(\bm{\eta})\triangleq \frac{\mathbb{E}_{\Omega}\{P_{\mathrm{C}}(\Omega, \bm{\eta})\}}{\mathbb{E}_{\Omega}\{P_{\mathrm{A}}(\Omega, \bm{\eta})\}} = \frac{\int_{-\Omega^{\prime}}^{\Omega^{\prime}}P_{\mathrm{C}}(\Omega, \bm{\eta})\mathrm{d}\Omega}{\int_{-\Omega^{\prime}}^{\Omega^{\prime}}P_{\mathrm{A}}(\Omega, \bm{\eta})\mathrm{d}\Omega}.
\end{equation}
Therefore, the window function that maximizes the average energy ratio can be obtained through the following optimization problem 
\begin{equation}\label{opt_win}
   \bm{\eta}^{\mathrm{o}} = \underset{\bm{\eta}\in\mathbb{C}^{M}}{\operatorname{argmax}}\ \lambda(\bm{\eta}).
\end{equation}

With the aim of solving the optimization problem in~\eqref{opt_win}, we define the matrices $\bm{\Phi}\in\mathbb{R}^{M\times M}$ and $\bm{\Xi}\in\mathbb{R}^{M\times M}$ with 
\begin{subequations}\label{Phi_Xi}
   \begin{align}
      [\bm{\Phi}]_{m, m^{\prime}} = &\mathrm{sinc}(\pi(m - m^{\prime})/S)D_{c}(2\pi(m - m^{\prime})/S),\\
      [\bm{\Xi}]_{m, m^{\prime}} = &\mathrm{sinc}(2\pi(m - m^{\prime})\Omega^{\prime}\Delta_{\mathrm{an}}^{-1}/S)\nonumber\\
      &\times D_{(A-1)/2}(2\pi(m - m^{\prime})/S),
   \end{align}
\end{subequations}
for $m,m^{\prime}\in\mathcal{Z}_{M}^{+}$, where 
\begin{subequations}
   \begin{align}
      \mathrm{sinc}(x)&\triangleq
      \left\{\begin{aligned}
      &\frac{\sin(x)}{x},&& x \neq 0\\
      &1,&& x = 0
      \end{aligned}\right.,\\
      D_{n}(x)&\triangleq
     \left\{\begin{aligned}
        &\frac{\sin((n+1/2)x)}{\sin(x/2)},&& x \neq 2k\pi, k\in\mathbb{Z}\\
        &2n+1,&& x = 2k\pi, k\in\mathbb{Z}
         \end{aligned}\right.,
   \end{align}
\end{subequations}
are the sinc function and the Dirichlet kernel, respectively. Then we have the following theorem and corollary, proved in Appendix~\ref{apd3} and Appendix~\ref{apd4}, respectively.

\begin{theorem}\label{theo2}
   The optimal window function of~\eqref{opt_win} is the generalized eigenvector associated with the maximum generalized eigenvalue of pair $\{\bm{\Phi}, \bm{\Xi}\}$.
\end{theorem}
\begin{corollary}\label{coro1}
   The optimal window function of~\eqref{opt_win} can be constructed as a real centrosymmetric vector, i.e., 
   \begin{equation}
      \bm{\eta}^{\mathrm{o}} = \overline{\mathbf{I}}_{M}\bm{\eta}^{\mathrm{o}}\in\mathbb{R}^{M},
   \end{equation}
   in which $\overline{\mathbf{I}}_{M}\triangleq [\mathbf{e}_{M}, \mathbf{e}_{M-1}, \cdots, \mathbf{e}_{1}]\in\mathbb{R}^{M\times M}$ is the reverse permutation matrix.
\end{corollary}

From Theorem~\ref{theo2} and Corollary~\ref{coro1}, the energy-focusing window, $\bm{\eta}^{\mathrm{o}}$, can be obtained by searching a real centrosymmetric-structured generalized eigenvector of a matrix pair, which can be computed offline and used for the windowed BSTR. 

\subsection{Efficient Implementation of Windowed BSTR}

In order to efficiently implement the windowed BSTR, we consider real centrosymmetric window functions, such as the proposed energy-focusing window $\bm{\eta}^{\mathrm{o}}$, the Hanning window, the Kaiser window, etc., and further discuss the low-complexity computations of $\widehat{\mathbf{y}}$, $\widehat{\mathbf{D}}$ and $\mathbf{U}$ in~\eqref{y_hat0} and~\eqref{D_hat} for $\widehat{\mathbf{y}}_{l}$, $\widehat{\mathbf{D}}_{l}$ and $\mathbf{U}_{l}$ in~\eqref{xhat_l_win} and~\eqref{Wl_hat}, in conjunction with their structural properties. 

The computation of $\widehat{\mathbf{y}}$ is required to be implemented at each turbo iteration since the a priori mean $\bm{\mu}$ is updated accordingly. Exploiting the structure of the beam matrix~\eqref{V_mtx} 
, we have 
\begin{align}\label{y_hat}
   \widehat{\mathbf{y}} =& \mathbf{V}^{\mathrm{H}}\bm{\Lambda}\left(\mathbf{y} - \mathbf{VG}\bm{\mu}\right)\nonumber\\
   =& -\frac{1}{M}\mathbf{I}_{A,S}\overline{\bm{\Pi}}_{S}^{(A-1)/2}\bm{\Omega}^{\ast}\mathbf{F}_{S}^{\mathrm{H}}\mathbf{I}_{S,M}\bm{\Lambda}\nonumber\\
   &\times\left(\sqrt{M}\mathbf{y} - \mathbf{I}_{M, S}\mathbf{F}_{S}\bm{\Omega}\overline{\bm{\Pi}}_{S}^{S - (A-1)/2}\mathbf{I}_{S, A}\mathbf{G}\bm{\mu}\right).
\end{align}
Then the computation of $\widehat{\mathbf{y}}$ can be efficiently implemented since $\mathbf{G}$ is sparse and FFT can be adopted. 

The computation of $\widehat{\mathbf{D}}$ is required to be implemented at each channel realization since it only varies with $\mathbf{G}$. By substituting ~\eqref{V_mtx}, $\widehat{\mathbf{D}}$ can be rewritten as $\widehat{\mathbf{D}} = \widehat{\mathbf{Q}}\mathbf{G} = \mathbf{I}_{A,S}\widetilde{\mathbf{Q}}\mathbf{I}_{S,A}\mathbf{G}$, where 
\begin{equation}\label{til_Q}
   \widetilde{\mathbf{Q}} = -\frac{1}{M}\overline{\bm{\Pi}}_{S}^{(A-1)/2}\bm{\Omega}^{\ast}\mathbf{F}_{S}^{\mathrm{H}}\mathbf{I}_{S, M}\bm{\Lambda}\mathbf{I}_{M, S}\mathbf{F}_{S}\bm{\Omega}\overline{\bm{\Pi}}_{S}^{S-(A-1)/2}.
\end{equation}
The following lemma, proved in Appendix~\ref{apd5}, demonstrates the structure of $\widetilde{\mathbf{Q}}$. 
\begin{lemma}\label{lemm2}
   When $\bm{\Lambda}$ is scaled to satisfy $\operatorname{tr}\{\bm{\Lambda}\} = M$, $\widetilde{\mathbf{Q}}$ is a real-valued Toeplitz matrix with the expression as follows:
   \begin{equation}\label{til_Q_sym}
      \widetilde{\mathbf{Q}} = \mathbf{I}_{S} + \sum_{k=1}^{\lceil S/2\rceil-1}\gamma_{k}\left(\overline{\bm{\Pi}}_{S}^{k} + (-1)^{M+1}\overline{\bm{\Pi}}_{S}^{S-k}\right),
   \end{equation}
   where $\gamma_{k}\triangleq \frac{1}{M}\mathbf{c}_{k}^{\mathrm{T}}\bm{\eta}\in\mathbb{R},\ k\in\mathbb{Z}$, and $[\mathbf{c}_{k}]_{m} = \cos(\pi k(M-2m+1)/S),\ k\in\mathbb{Z},m\in\mathcal{Z}_{M}^{+}$.
\end{lemma}

\begin{figure}[ht]
	\centering
	\includegraphics[width=0.48\textwidth]{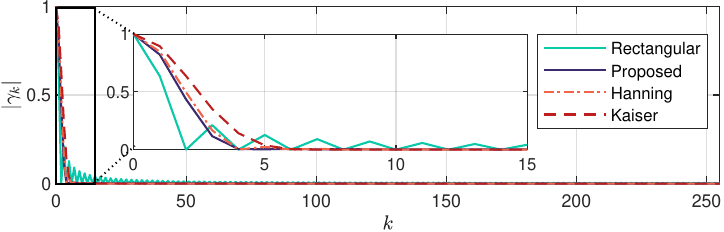}
	\caption{Magnitude of $\gamma_{k}$ under different window functions with $M=256$, $F=2$. For the proposed window $\mathbf{\eta}^{\mathrm{o}}$, $c = 3$, $\Omega^{\prime} = 1$. The shape parameter of the Kaiser window is set as 10.}
	\label{fig:gamma}
\end{figure}

As illustrated in Fig.~\ref{fig:gamma}, when we adopt the proposed energy-focusing window $\bm{\eta}^{\mathrm{o}}$ or some classical windows, there are numerous extremely small values of $\gamma_{k},\ k \in\mathcal{Z}_{\lceil S/2\rceil-1}^{+}$. To reduce the computational complexity of $\widehat{\mathbf{D}}$, we make an approximation of $\widetilde{\mathbf{Q}}$ by omitting $\gamma_{k},\ k\notin\mathcal{Q}$, where the filtering set $\mathcal{Q} \triangleq \left\{k\in\mathcal{Z}_{\lceil S/2\rceil-1}^{+} \big| |\gamma_{k}| > \varepsilon \right\}$ with $Q = |\mathcal{Q}|$
and $\varepsilon$ is a threshold. Then $\widetilde{\mathbf{Q}} \approx \mathbf{I}_{S} + \sum_{k\in\mathcal{Q}}\gamma_{k}\big(\overline{\bm{\Pi}}_{S}^{k} + (-1)^{M+1}\overline{\bm{\Pi}}_{S}^{S-k}\big)$, thus $\widehat{\mathbf{D}}$ can be approximated as 
\begin{align}\label{D_hat_appro}
   \widehat{\mathbf{D}}&=\mathbf{I}_{A,S}\widetilde{\mathbf{Q}}\mathbf{I}_{S,A}\mathbf{G}\nonumber\\
   &\approx \mathbf{G} + \underset{\triangleq\mathbf{G}_{\mathrm{d}}}{\underbrace{\mathbf{I}_{A,S}\sum_{k\in\mathcal{Q}}\left(\overline{\bm{\Pi}}_{S}^{k} \!+\! (\!-1)^{M+1}\overline{\bm{\Pi}}_{S}^{S-k}\right)\mathbf{I}_{S,A}\mathbf{G}\gamma_{k}}},
\end{align}
where $\mathbf{G}_{\mathrm{d}}$ can be viewed as diffusion of beam domain channel $\mathbf{G}$.
From~\eqref{D_hat_appro}, the approximate computation of $\widehat{\mathbf{D}}$ arises from the computation of $\mathbf{G}_{\mathrm{d}}$, involving the multiplications of a sparse complex-valued matrix and real-valued scalars (i.e., $\mathbf{G}\gamma_{k},\ k\in\mathcal{Q}$), along with several permutations, sign inversions, extractions, and summation operations. 

The computation of $\mathbf{U}$ can be realized offline. Similar to~\eqref{til_Q} and~\eqref{til_Q_sym}, $\mathbf{U}$ is also a real matrix with the structure  
\begin{align}\label{U}
   \mathbf{U} =& -\frac{1}{M}\mathbf{I}_{A,S}\overline{\bm{\Pi}}_{S}^{(A-1)/2}\bm{\Omega}^{\ast}\mathbf{F}_{S}^{\mathrm{H}}\mathbf{I}_{S, M}\bm{\Lambda}^{2}\nonumber\\
   &\times\mathbf{I}_{M, S}\mathbf{F}_{S}\bm{\Omega}\overline{\bm{\Pi}}_{S}^{S-(A-1)/2}\mathbf{I}_{S,A}\nonumber\\
   =& \widetilde{\gamma}_{0}\mathbf{I}_{A} \!+ \!\!\!\!\sum_{k=1}^{\lceil S/2\rceil-1}\!\!\!\!\widetilde{\gamma}_{k}\mathbf{I}_{A, S}\left(\overline{\bm{\Pi}}_{S}^{k} \!+\! (-1)^{M+1}\overline{\bm{\Pi}}_{S}^{S-k}\right)\mathbf{I}_{S, A},
\end{align}
where $\widetilde{\gamma}_{k}\triangleq \frac{1}{M}\mathbf{c}_{k}^{\mathrm{T}}(\bm{\eta}\odot \bm{\eta})\in\mathbb{R},\ k\in\mathbb{Z}$. Eq.~\eqref{U} indicates that $\mathbf{U}$ can be sufficiently characterized by $\lceil S/2\rceil$ real-valued scalars $\widetilde{\gamma}_{k},\ k\in\mathcal{Z}_{\lceil S/2\rceil}^{+}$. Consequently, the required storage space and computational complexity are relatively low.

As a special case, the computation of $\widetilde{\mathbf{y}}$, $\mathbf{D}$ and $\mathbf{Q}$ in~\eqref{y_til} and~\eqref{D_Q} of BSTR are respectively consistent with~\eqref{y_hat},~\eqref{D_hat_appro} and~\eqref{U} by removing the window function $\bm{\Lambda}$, in which case $\gamma_{k} = \widetilde{\gamma}_{k} = \sin(\pi k/F) / (M\sin(\pi k/S)),\ k\in\mathbb{Z}$.

Recalling the intention of window design, the inter-group interference can be effectively suppressed by adopting the proper window function in windowed BSTR. Upon this, we define the interference UT set for group $l$
\begin{equation}
   \widetilde{\mathcal{N}}_{l}\triangleq \left\{u\in\mathcal{Z}_{U}^{+}\big| \mathcal{A}_{u}\cap\mathcal{B}_{l}\neq\varnothing \right\}\ \text{with}\ \widetilde{N}_{l} = \big|\widetilde{\mathcal{N}}_{l}\big|.
\end{equation}
Then $\widehat{\mathbf{W}}_{l}$ in~\eqref{Wl_hat} can be approximated as 
\begin{align}\label{W_approx}
   \widehat{\mathbf{W}}_{l}&\approx\left(\underline{\mathbf{D}}_{l}\underline{\bm{\Sigma}}_{l}\underline{\mathbf{D}}_{l}^{\mathrm{H}} + \sigma_{\mathrm{z}}\mathbf{U}_{l}\right)^{-1}\overline{\mathbf{D}}_{l}\bm{\Sigma}_{l}\nonumber\\
   &= \mathbf{T}_{l}\big(\underline{\bm{\Sigma}}_{l}\mathbf{K}_{l} + \sigma_{\mathrm{z}}\mathbf{I}\big)^{-1}\mathbf{S}_{l}\bm{\Sigma}_{l},
\end{align}
where $\underline{\mathbf{D}}_{l}\triangleq \widehat{\mathbf{D}}_{l}\widetilde{\mathbf{N}}_{l}\in\mathbb{C}^{B_{l}\times\widetilde{N}_{l}}$ with $\widetilde{\mathbf{N}}_{l}\triangleq \mathbf{S}\big(U, \widetilde{\mathcal{N}}_{l}\big)\in\mathbb{R}^{U\times\widetilde{N}_{l}}$, $\mathbf{S}_{l}\triangleq \widetilde{\mathbf{N}}_{l}^{\mathrm{T}}\mathbf{N}_{l}\in\mathbb{R}^{\widetilde{N}_{l}\times N_{l}}$ is a selection matrix, $\underline{\bm{\Sigma}}_{l}\triangleq \widetilde{\mathbf{N}}_{l}^{\mathrm{T}}\bm{\Sigma}\widetilde{\mathbf{N}}_{l}\in\mathbb{R}^{\widetilde{N}_{l}\times \widetilde{N}_{l}}$, and
\begin{subequations}\label{Tl_Kl}
   \begin{align}
      \mathbf{T}_{l}&\triangleq \mathbf{U}_{l}^{-1}\underline{\mathbf{D}}_{l}\in\mathbb{C}^{B_{l}\times\widetilde{N}_{l}},\\
      \mathbf{K}_{l}&\triangleq \underline{\mathbf{D}}_{l}^{\mathrm{H}}\mathbf{T}_{l}\in\mathbb{C}^{\widetilde{N}_{l}\times\widetilde{N}_{l}}.
   \end{align}
\end{subequations}
Since matrix $\mathbf{U}$ is fixed, $\mathbf{U}_{l}^{-1}$ can be precomputed. Additionally, $\mathbf{T}_{l}$ and $\mathbf{K}_{l}$ can be computed before turbo iterations, as they are independent of $\bm{\mu}$ and $\bm{\Sigma}$. At each iteration, the computational complexity of the inversion term $\big(\underline{\bm{\Sigma}}_{l}\mathbf{K}_{l} + \sigma_{\mathrm{z}}\mathbf{I}\big)^{-1}$ in~\eqref{W_approx} is much lower than that in~\eqref{Wl_hat}, due to the size $\widetilde{N}_{l}$ of interference UT set is typically smaller than the size $B_{l}$ of the beam set. 

In summary, combined with the low-complexity computation strategy in~\eqref{y_hat} and the approximations in~\eqref{D_hat_appro} and~\eqref{W_approx}, the computational complexity of the windowed BSTR can be remarkably reduced.

\subsection{Complexity Analysis}
In this subsection, we evaluate the computational complexity of the proposed receivers in more details by counting the number of complex multiplications (CMs)~\cite{7163554}, where the multiplication between a real scalar and a complex scalar is viewed as half CM. To better examine the computational complexity of turbo receivers, only the complexity of signal detection is considered since it is dominant, while the complexities of other modulos (e.g., the extrinsic information calculation, the a priori information update and decoding) are not taken into consideration. The number of iterations is denoted as $T$.
Average values $\widetilde{A}\triangleq\frac{1}{U}\sum_{u=1}^{U}A_{u}$, $B\triangleq\frac{1}{L}\sum_{l=1}^{L}B_{l}$, $N\triangleq\frac{1}{L}\sum_{l=1}^{L}N_{l}$, and $\widetilde{N}\triangleq\frac{1}{L}\sum_{l=1}^{L}\widetilde{N}_{l}$ are defined for the convenience of complexity analysis.  

To begin with, we analyze the computational complexity of the BSTR. Before the iteration, $\mathbf{D}$ needs to be computed, which requires $\widetilde{A}U(\lceil S/2\rceil - 1)/2$ CMs by combining the structure~\eqref{D_hat_appro} with $\bm{\Lambda} = \mathbf{I}_{M}$ and $\epsilon = 0$. The complexity for each iteration comes from the computation of $\widehat{\mathbf{x}}_{l},\ l\in\mathcal{Z}_{L}^{+}$, where the computation of $\widetilde{\mathbf{y}}$ can be efficiently implemented by~\eqref{y_hat} with $\widetilde{A}U + A + S(1+\log S)$ CMs, and the computation of~\eqref{xhat_l} consumes $(B(B+2)U + B^2(B+3))L/2 + B(B+1)U$ CMs. Therefore, the total complexity of the BSTR is $\widetilde{A}U(\lceil S/2\rceil - 1)/2 + (\widetilde{A}U + A + S(1+\log S) + (B(B+2)U + B^2(B+3))L/2 + B(B+1)U)T$.

For windowed BSTR, the computation of $\widehat{\mathbf{D}}$ requires $\widetilde{A}UQ/2$ CMs by approximation~\eqref{D_hat_appro}. Before the iteration, $\mathbf{T}_{l}$ and $\mathbf{K}_{l}$ in~\eqref{Tl_Kl} are also needed to be computed, which needs $(B+(\widetilde{N} + 1)/2)B\widetilde{N}L$ CMs. The complexity for each iteration stems from the computation of $\widehat{\mathbf{x}}_{l}^{\mathrm{win}},\ l\in\mathcal{Z}_{L}^{+}$, involving $\widetilde{A}U + A + M/2 + S(1+\log S) + (\widetilde{N}/2+2)\widetilde{N}^{2}L + (\widetilde{N}/2 + B\widetilde{N}+B)U$ CMs. Thus the total complexity of the windowed BSTR is $\widetilde{A}UQ/2 + (B+(\widetilde{N} + 1)/2)B\widetilde{N}L + (\widetilde{A}U + A + M/2 + S(1+\log S) + (\widetilde{N}/2+2)\widetilde{N}^{2}L + (\widetilde{N}/2 + B\widetilde{N}+B)U)T$.

The computational complexity of the MMSE turbo receiver is also included as a benchmark. Before the iteration, $\mathbf{H}^{\mathrm{H}}\mathbf{H}$ is computed with $MU(U+1)/2$ CMs. At each iteration, the complexity of signal detection is given by $(U+5)U^{2}/2 + MU(U+2)$. Then the complexity of the MMSE turbo receiver is $MU(U+1)/2 + ((U+5)U^{2}/2 + MU(U+2))T$ in total. In massive MIMO systems, when $M$ and $U$ are both large, the proposed BSTRs have a markedly lower complexity in comparison to the MMSE turbo receiver.

\section{Simulation Results}\label{Sim}
In this section, simulation results of the proposed receivers for HF skywave massive MIMO systems are provided. 

The simulation parameters are presented in Table~\ref{tab1}. UTs are distributed in a $140^{\circ}$ sector around $2000$ km from the BS with $L = 18$ groups, and UT scheduling is performed to ensure that UTs from different groups exhibit relatively low channel correlation. Channel parameters are generated by the ray-tracing software Proplab-Pro 3.1~\cite{10122719}, including the signal strength and the propagation distance of each UT, etc. The BS and UTs are equipped with isotropic antennas. Accordingly, the space domain channel, $\mathbf{H}$, can be obtained, and the corresponding beam domain channel can be acquired by the approach proposed in~\cite{HFmimo}. The modulation scheme is quadrature amplitude modulation (QAM), the channel code is selected as low density parity check (LDPC) code of length $N_{\mathrm{c}} = 2112$ with code rate $3/4$, and a row-column interleaver with length $N_{\mathrm{c}}/N$ is adopted. 

We adopt the bit-error rate (BER) as the performance metric, and the signal-to-noise ratio (SNR) refers to the received SNR. For the proposed energy-focusing window, $\bm{\eta}^{\mathrm{o}}$, we set $c=3$ and $\Omega^{\prime} = 1$. Specifically, the following receivers are compared:   
\begin{itemize}
   \item \textbf{BSTR:} Beam structured turbo receiver, proposed in section~\ref{BSTR}.
   \item \textbf{WBSTR:} Windowed BSTR, proposed in section~\ref{WBSTR}.
   \item \textbf{MMSE TR:} MMSE turbo receiver~\cite{984761}.
   \item \textbf{BSD:} Beam structured detector~\cite{BSD}.
   \item \textbf{JSTD:} Joint Slepian transform based detector~\cite{10122719}.
\end{itemize}
\begin{table}[htbp]
	\renewcommand\arraystretch{1}
	\caption{Simulation Parameters}
	\centering
	\setlength{\tabcolsep}{5pt}
	\begin{tabular}{cc}
      \hline
		 Parameter & Value  \\ 
      \hline
		 Carrier frequency $f_{\mathrm{c}}$ & $16$ MHz \\
       Number of antennas at the BS $M$ & $256$\\
       Antenna spacing at the BS $d$ & $9$ m\\
       Fine factor $F$ & $2$ \\
       Number of UTs $U$ & $72$\\
      \hline	
	\end{tabular}
	\label{tab1}
\end{table}

\begin{figure}[htbp]
   \centering
    \subfigure[]{
    \includegraphics[width=7.7cm]{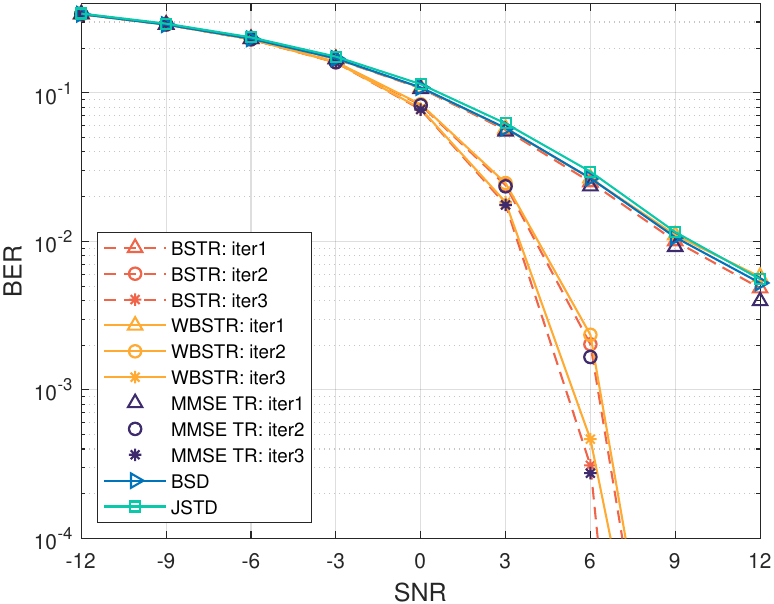}
    \label{ber_qpsk}
    }
    \subfigure[]{
      \includegraphics[width=7.7cm]{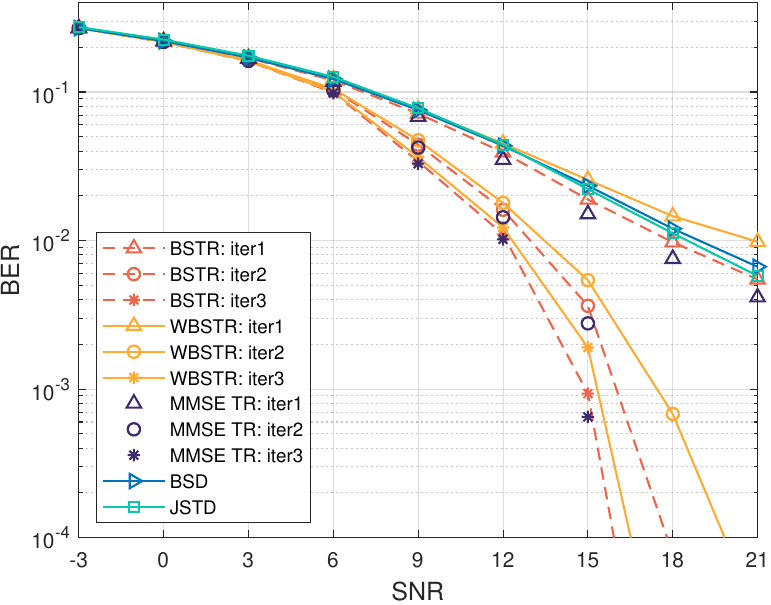}
      \label{ber_16qam}
    }
    \caption{BER performance of the BSTR and the windowed BSTR, compared with other receivers. (a) $4$-QAM; (b) $16$-QAM.}
    \label{diff_mod}
\end{figure}

Fig.~\ref{diff_mod} provides the BER performance of the BSTR and the windowed BSTR under different modulation orders, compared with the MMSE TR, BSD and JSTD. The number of turbo iterations is set as $3$, the proposed energy-focusing window, $\bm{\eta}^{\mathrm{o}}$, is adopted and the threshold for the windowed BSTR is set as $\epsilon = 10^{-3}$. For 4-QAM modulation, the performance of the proposed receivers is approaching that of the MMSE TR at each iteration. For 16-QAM modulation, the BSTR outperforms the BSD and JSTD at the first iteration, and the performance gap is enlarged during the subsequent iterations. The performance of the windowed BSTR is inferior to that of the BSTR in the first few iterations, especially in high SNR region. This phenomenon can be attributed to the approximations implemented by the windowed BSTR in the detection process. This effect is particularly pronounced in high SNR scenarios, where interference predominates, leading to a more significant impact of these approximations on performance. However, for windowed BSTR, these approximations result in a substantial reduction in computational complexity while iterative processes can provide significant performance gain and the performance gap with the BSTR and the MMSE TR is extremely small at the final iteration. After the first iteration, the performance of the windowed BSTR is much better than the first iteration of the MMSE TR (i.e., the MMSE detection without the a priori information update).

\begin{figure}[htbp] 
   \centering
    \subfigure[]{
    \includegraphics[width=7.7cm]{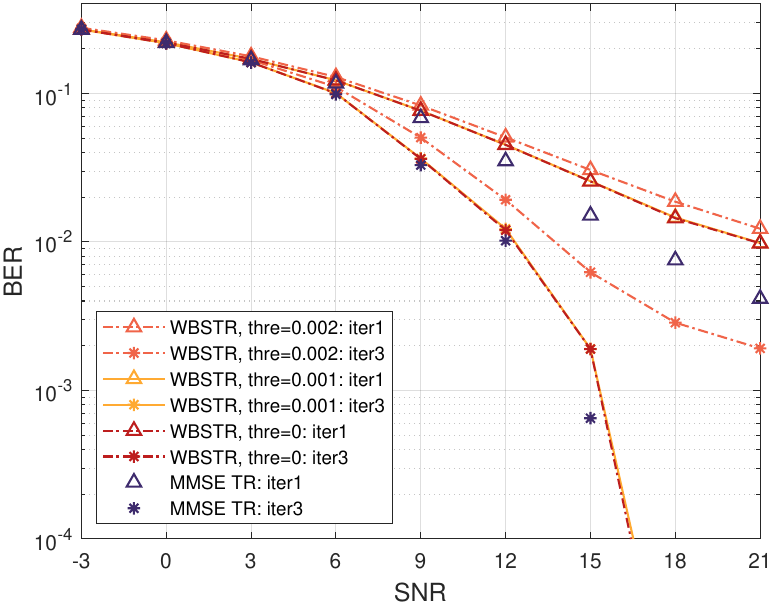}
    \label{ber_diffthre}
    }
    \subfigure[]{
      \includegraphics[width=7.7cm]{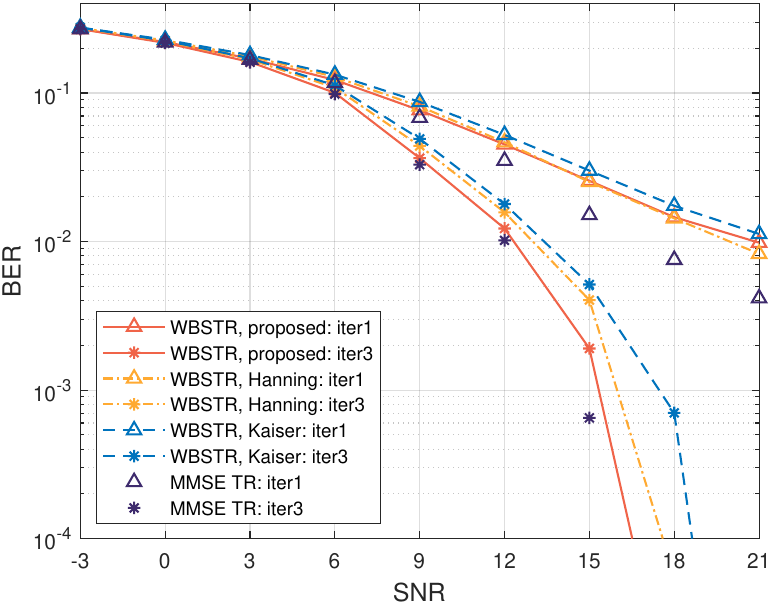}
      \label{ber_diffwin}
    }
    \caption{BER performance of the windowed BSTR with (a) different threshold $\epsilon$ for energy-focusing window (b) different window function.}
    \label{diff_win}
\end{figure}

Fig.~\ref{diff_win} illustrates the BER performance of the windowed BSTR under different parameters and window functions, including the MMSE TR as baseline, where the modulation scheme is $16$-QAM. In Fig.~\ref{ber_diffthre}, the performances of the windowed BSTR with $\bm{\eta}^{\mathrm{o}}$ against various thresholds $\epsilon$ are compared. It can be observed that the performance with $\epsilon = 10^{-3}$ ($Q=27$) is almost same as the performance with $\epsilon = 0$ ($Q = 255$, i.e., without approximate computation of $\widehat{\mathbf{D}}$), while it suffers severe performance degradation when $\epsilon = 2\times 10^{-3}$ ($Q=14$). Since the computational complexity of $\widehat{\mathbf{D}}$ is proportional to $Q$, then $\epsilon = 10^{-3}$ is a preferable choice that achieves satisfactory performance while maintaining low complexity. Fig.~\ref{ber_diffwin} compares the performances of the windowed BSTR with different window functions, including the proposed energy-focusing window, the Hanning window, and the Kaiser window, where the shape parameter of the Kaiser window is 10. For the sake of fairness, the threshold $\epsilon$ for each window function is set to satisfy $Q=27$, which ensures that the computational complexity of the windowed BSTR remains equal for each window function. From the figure, the performance of the proposed energy-focusing window is better than all the other window functions at the final iteration. Moreover, during the final iteration, the windowed BSTR with all considered window functions outperforms the first iteration of the MMSE TR. This conforms the effectiveness of the windowed BSTR and the proposed energy-focusing window.



\begin{figure}[ht]
	\centering
	\includegraphics[width=7.7cm]{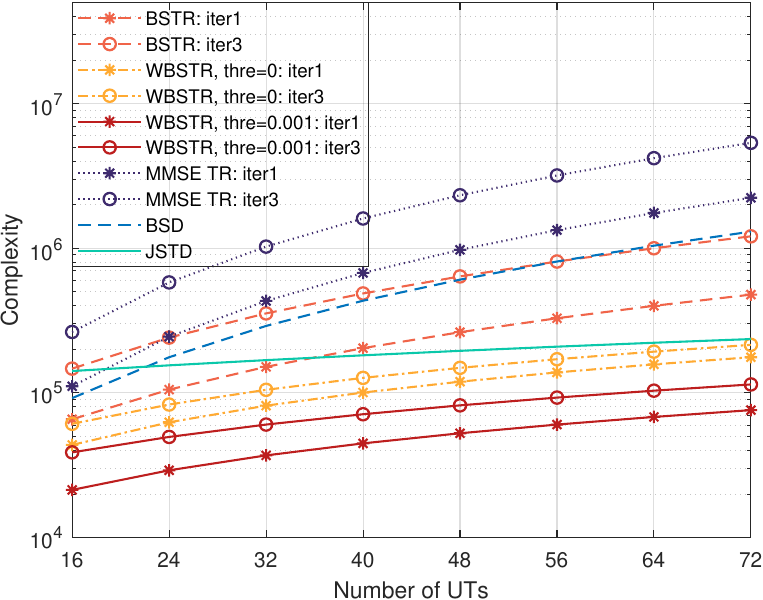}
	\caption {Computational complexity of the BSTR and the windowed BSTR, compared with other receivers.}
	\label{fig:complexity}
\end{figure}

In Fig.~\ref{fig:complexity}, the computational complexities of the BSTR and the windowed BSTR are compared versus the number of UTs, together with the MMSE TR, the JSTD and the BSD. The proposed energy-focusing window is adopted. From the figure, the complexity of the BSTR with three iterations is close to that of the BSD and lower than that of the MMSE TR with one iteration, while it is higher than that of the JSTD when the number of UT is large. For the windowed BSTR, due to the approximation in~\eqref{W_approx}, the complexity after even three iterations is lower than that of all the other receivers, regardless of whether $\epsilon = 10^{-3}$ or $\epsilon = 0$. Furthermore, owning to the approximation in~\eqref{D_hat_appro}, the complexity of the windowed BSTR with $\epsilon = 10^{-3}$ is obviously lower than that of $\epsilon = 0$, while the BER performance stays consistent, as illustrated in Fig.~\ref{ber_diffthre}. Therefore, compared to the one-time MMSE TR, the BSD and the JSTD, the windowed BSTR achieves a superior performance with a significantly reduced complexity. 

\section{Conclusions}\label{Con}
In this paper, we investigated receiver design for HF skywave massive MIMO systems. We first proposed the modified BBCM and constructed the beam matrix as DFT-based structure for low-complexity receiver design. Then, we proposed the BSTR, involving low-dimensional beam domain signal detection
for each UT group. We proved that for BSTR, the beam structured signal detection is asymptotically optimal in terms of minimizing MSE when the number of BS antennas is sufficiently large. To further reduce the computational complexity, we proposed the windowed BSTR by incorporating a window function in BSTR. A well-designed energy-focusing window was introduced for windowed BSTR. The windowed BSTR was efficiently implemented by using the characteristics of the beam matrix and the beam domain channel sparsity. Finally, the effectiveness of the proposed receivers was demonstrated by simulation results.

\appendices

\section{Proof of Theorem~\ref{theo1}}\label{apd2}
Define $\mathcal{B}\triangleq\cup_{l\in\mathcal{Z}_{L}^{+}}\mathcal{B}_{l}$ and $\mathbf{B} \triangleq \mathbf{S}(A,\mathcal{B})\in\mathbb{R}^{A\times B}$ with $B = |\mathcal{B}|$. Then the modified BBCM can be rewritten as $\mathbf{H} = \widetilde{\mathbf{V}}\widetilde{\mathbf{G}}$, where $\widetilde{\mathbf{V}}\triangleq\mathbf{VB}\in\mathbb{C}^{M\times B}$ and $\widetilde{\mathbf{G}}\triangleq\mathbf{B}^{\mathrm{T}}\mathbf{G}\in\mathbb{C}^{B\times U}$. Combined with the condition of minimizing MSE $\frac{\partial\epsilon_{l}}{\partial\mathbf{n}_{l}} = \mathbf{0}$, we have $\mathbf{n}_{l} = \bm{\mu}_{l} - \widetilde{\mathbf{W}}_{l}^{\mathrm{H}}\widetilde{\mathbf{G}}\bm{\mu}$, where $\widetilde{\mathbf{W}}_{l} = \widetilde{\mathbf{V}}^{\mathrm{H}}\mathbf{R}_{l}$, and the MSE function can be expressed as 
\begin{align}
   \epsilon_{l} =& \operatorname{tr}\Big\{\widetilde{\mathbf{W}}_{l}^{\mathrm{H}}\widetilde{\mathbf{G}}\bm{\Sigma}\widetilde{\mathbf{G}}^{\mathrm{H}}\widetilde{\mathbf{W}}_{l} + \sigma_{\mathrm{z}}\mathbf{R}_{l}^{\mathrm{H}}\mathbf{R}_{l} - \bm{\Sigma}_{l}\mathbf{N}_{l}^{\mathrm{T}}\widetilde{\mathbf{G}}^{\mathrm{H}}\widetilde{\mathbf{W}}_{l} \nonumber\\
   &- \widetilde{\mathbf{W}}_{l}^{\mathrm{H}}\widetilde{\mathbf{G}}\mathbf{N}_{l}\bm{\Sigma}_{l} + \bm{\Sigma}_{l}\Big\},
\end{align}
where $\widetilde{\mathbf{W}}_{l} = \widetilde{\mathbf{V}}^{\mathrm{H}}\mathbf{R}_{l}$. We turn to expressing $\epsilon_{l}$ as a univariate function of $\widetilde{\mathbf{W}}_{l}$. To ensure the existence of $\mathbf{R}_{l}$ when $\widetilde{\mathbf{W}}_{l}$ is given, $\widetilde{\mathbf{W}}_{l}$ should satisfy $\big(\widetilde{\mathbf{V}}^{\mathrm{H}}\big(\widetilde{\mathbf{V}}^{\mathrm{H}}\big)^{\dagger} - \mathbf{I}\big)\widetilde{\mathbf{W}}_{l} = \mathbf{O}$~\cite{penrose1955generalized}. Moreover, for a given optimal solution $\widetilde{\mathbf{W}}_{l}^{\mathrm{op}} = \widetilde{\mathbf{V}}^{\mathrm{H}}\mathbf{R}_{l}^{\mathrm{op}}$ of minimizing $\epsilon_{l}$, $\operatorname{tr}\big\{(\mathbf{R}_{l}^{\mathrm{op}})^{\mathrm{H}}\mathbf{R}_{l}^{\mathrm{op}}\big\}$ should be minimal. Otherwise, a smaller value of $\operatorname{tr}\big\{(\mathbf{R}_{l}^{\mathrm{op}})^{\mathrm{H}}\mathbf{R}_{l}^{\mathrm{op}}\big\}$ leads to a lower MSE, which contradicts the optimality. Therefore, $\mathbf{R}_{l}^{\mathrm{op}}$ can be obtained through the following optimization problem
\begin{equation}\label{Rl_minpow}
   \mathbf{R}_{l}^{\mathrm{op}} = \underset{\mathbf{R}_{l}\in\mathbb{C}^{M\times N_{l}}}{\operatorname{argmin}} \operatorname{tr}\big\{\mathbf{R}_{l}^{\mathrm{H}}\mathbf{R}_{l}\big\},\  \text{s.t.}\ \widetilde{\mathbf{V}}^{\mathrm{H}}\mathbf{R}_{l} = \widetilde{\mathbf{W}}_{l}^{\mathrm{op}}.
\end{equation}
The problem~\eqref{Rl_minpow} can be equivalently expressed as 
\begin{align}\label{R_l_op}
   \operatorname{vec}\{\mathbf{R}_{l}^{\mathrm{op}}\} &= \underset{\operatorname{vec}\{\mathbf{R}_{l}\}\in\mathbb{C}^{M N_{l}}}{\operatorname{argmin}} \operatorname{vec}\{\mathbf{R}_{l}\}^{\mathrm{H}}\operatorname{vec}\{\mathbf{R}_{l}\},\nonumber\\
   &\text{s.t.}\ \big(\mathbf{I}\otimes\widetilde{\mathbf{V}}^{\mathrm{H}}\big)\operatorname{vec}\{\mathbf{R}_{l}\} = \operatorname{vec}\big\{\widetilde{\mathbf{W}}_{l}^{\mathrm{op}}\big\},
\end{align}
where $\operatorname{vec}\{\cdot\}$ denotes the vectorizing operation, and the optimal solution of~\eqref{R_l_op} can be obtained as~\cite{10419346} 
\begin{equation}
   \operatorname{vec}\{\mathbf{R}_{l}^{\mathrm{op}}\} = \left(\mathbf{I}\otimes \widetilde{\mathbf{V}}^{\mathrm{H}}\big(\widetilde{\mathbf{V}}^{\mathrm{H}}\widetilde{\mathbf{V}}\big)^{\dagger}\right)\operatorname{vec}\big\{\widetilde{\mathbf{W}}_{l}^{\mathrm{op}}\big\},
\end{equation}
thus $\mathbf{R}_{l}^{\mathrm{op}} = \widetilde{\mathbf{V}}^{\mathrm{H}}\big(\widetilde{\mathbf{V}}^{\mathrm{H}}\widetilde{\mathbf{V}}\big)^{\dagger}\widetilde{\mathbf{W}}_{l}^{\mathrm{op}}$. Then $\widetilde{\mathbf{W}}_{l}^{\mathrm{op}}$ can be obtained by solving the following optimization problem 
\begin{align}\label{W_til_opt}
   \widetilde{\mathbf{W}}_{l}^{\mathrm{op}} =& \underset{\widetilde{\mathbf{W}}_{l}\in\mathbb{C}^{B\times N_{l}}}{\operatorname{argmin}} \epsilon_{l}\big(\widetilde{\mathbf{W}}_{l}\big),\nonumber\\
   \text{s.t.}&\ \big(\widetilde{\mathbf{V}}^{\mathrm{H}}\big(\widetilde{\mathbf{V}}^{\mathrm{H}}\big)^{\dagger} - \mathbf{I}\big)\widetilde{\mathbf{W}}_{l} = \mathbf{O},
\end{align}
where the MSE function 
\begin{align}\label{epsl}
   \epsilon_{l}\big(\widetilde{\mathbf{W}}_{l}\big) &= \operatorname{tr}\Big\{\widetilde{\mathbf{W}}_{l}^{\mathrm{H}}\big(\widetilde{\mathbf{G}}\bm{\Sigma}\widetilde{\mathbf{G}}^{\mathrm{H}} + \sigma_{\mathrm{z}}\big(\widetilde{\mathbf{V}}^{\mathrm{H}}\widetilde{\mathbf{V}}\big)^{\dagger}\big)\widetilde{\mathbf{W}}_{l}\nonumber\\ 
   &- \bm{\Sigma}_{l}\mathbf{N}_{l}^{\mathrm{T}}\widetilde{\mathbf{G}}^{\mathrm{H}}\widetilde{\mathbf{W}}_{l} - \widetilde{\mathbf{W}}_{l}^{\mathrm{H}}\widetilde{\mathbf{G}}\mathbf{N}_{l}\bm{\Sigma}_{l} + \bm{\Sigma}_{l}\Big\}.
\end{align}
When $\widetilde{\mathbf{W}}_{l}^{\mathrm{op}}$ is obtained by solving~\eqref{W_til_opt}, the optimal detector can be obtained as $\mathbf{R}_{l}^{\mathrm{op}} = \widetilde{\mathbf{V}}^{\mathrm{H}}\big(\widetilde{\mathbf{V}}^{\mathrm{H}}\widetilde{\mathbf{V}}\big)^{\dagger}\widetilde{\mathbf{W}}_{l}^{\mathrm{op}}$ and $\mathbf{n}_{l}^{\mathrm{op}} = \bm{\mu}_{l} - \big(\widetilde{\mathbf{W}}_{l}^{\mathrm{op}}\big)^{\mathrm{H}}\widetilde{\mathbf{G}}\bm{\mu}$.

Since the UT number is limited and the directional cosines of the paths from different UTs are distinct, it can be checked that $\lim_{M\rightarrow\infty}\widetilde{\mathbf{V}}^{\mathrm{H}}\big(\widetilde{\mathbf{V}}^{\mathrm{H}}\big)^{\dagger} = \lim_{M\rightarrow\infty}\big(\widetilde{\mathbf{V}}^{\mathrm{H}}\widetilde{\mathbf{V}}\big)^{\dagger} = \mathbf{I}$~\cite{yu2021hf}. Thus when $M\rightarrow\infty$, the optimal detector can be obtained as 
\begin{subequations}\label{Rl_dl_op_inf}
   \begin{align}
      \mathbf{R}_{l}^{\mathrm{op}} &= \widetilde{\mathbf{V}}^{\mathrm{H}}\widetilde{\mathbf{W}}_{l}^{\mathrm{op}},\\
      \mathbf{n}_{l}^{\mathrm{op}} &= \bm{\mu}_{l} - \big(\widetilde{\mathbf{W}}_{l}^{\mathrm{op}}\big)^{\mathrm{H}}\widetilde{\mathbf{G}}\bm{\mu}.
   \end{align}
\end{subequations}
where $\widetilde{\mathbf{W}}_{l}^{\mathrm{op}} = \operatorname{argmin}_{\widetilde{\mathbf{W}}_{l}\in\mathbb{C}^{B\times N_{l}}} \epsilon_{l}\big(\widetilde{\mathbf{W}}_{l}\big)$,
with 
\begin{align}
   \epsilon_{l}\big(\widetilde{\mathbf{W}}_{l}\big) =& \operatorname{tr}\Big\{\widetilde{\mathbf{W}}_{l}^{\mathrm{H}}\big(\widetilde{\mathbf{G}}\bm{\Sigma}\widetilde{\mathbf{G}}^{\mathrm{H}} + \sigma_{\mathrm{z}}\mathbf{I}\big)\widetilde{\mathbf{W}}_{l}\nonumber\\ 
   &- \bm{\Sigma}_{l}\mathbf{N}_{l}^{\mathrm{T}}\widetilde{\mathbf{G}}^{\mathrm{H}}\widetilde{\mathbf{W}}_{l} - \widetilde{\mathbf{W}}_{l}^{\mathrm{H}}\widetilde{\mathbf{G}}\mathbf{N}_{l}\bm{\Sigma}_{l} + \bm{\Sigma}_{l}\Big\}.
\end{align}

We further denote $\overline{\mathbf{W}}_{l} = \mathbf{B}^{\mathrm{T}}\mathbf{B}_{l}\mathbf{B}_{l}^{\mathrm{T}}\mathbf{B}\widetilde{\mathbf{W}}_{l}^{\mathrm{op}}$, i.e., the values corresponding to the non-zero beam positions of group $l$ within $\widetilde{\mathbf{W}}_{l}^{\mathrm{op}}$ are preserved, while the values at other positions are set to zero. Then, 
\begin{equation}\label{Wbar_geq}
   \epsilon_{l}\big(\overline{\mathbf{W}}_{l}\big)\geq \epsilon_{l}\big(\widetilde{\mathbf{W}}_{l}^{\mathrm{op}}\big).
\end{equation}
On the other hand, we have $\operatorname{tr}\Big\{\overline{\mathbf{W}}_{l}^{\mathrm{H}}\big(\widetilde{\mathbf{G}}\bm{\Sigma}\widetilde{\mathbf{G}}^{\mathrm{H}} + \sigma_{\mathrm{z}}\mathbf{I}\big)\overline{\mathbf{W}}_{l}\Big\} \leq \operatorname{tr}\Big\{\big(\widetilde{\mathbf{W}}_{l}^{\mathrm{op}}\big)^{\mathrm{H}}\big(\widetilde{\mathbf{G}}\bm{\Sigma}\widetilde{\mathbf{G}}^{\mathrm{H}} + \sigma_{\mathrm{z}}\mathbf{I}\big)\widetilde{\mathbf{W}}_{l}^{\mathrm{op}}\Big\}$ when $\forall l\neq l^{\prime},\ \mathcal{B}_{l}\cap\mathcal{B}_{l^{\prime}} = \varnothing$, and $\mathbf{N}_{l}^{\mathrm{T}}\widetilde{\mathbf{G}}^{\mathrm{H}}\overline{\mathbf{W}}_{l} = \mathbf{N}_{l}^{\mathrm{T}}\widetilde{\mathbf{G}}^{\mathrm{H}}\widetilde{\mathbf{W}}_{l}^{\mathrm{op}}$, which leads to 
\begin{equation}\label{Wbar_leq}
   \epsilon_{l}\big(\overline{\mathbf{W}}_{l}\big)\leq \epsilon_{l}\big(\widetilde{\mathbf{W}}_{l}^{\mathrm{op}}\big).
\end{equation}
Combining~\eqref{Wbar_geq} and~\eqref{Wbar_leq} yields $\epsilon_{l}\big(\overline{\mathbf{W}}_{l}\big) = \epsilon_{l}\big(\widetilde{\mathbf{W}}_{l}^{\mathrm{op}}\big)$, which means $\overline{\mathbf{W}}_{l} = \widetilde{\mathbf{W}}_{l}^{\mathrm{op}}$. This implies the fact that, without sacrificing the optimality, we can focus on designing values in the non-zero beam positions of group $l$ within $\widetilde{\mathbf{W}}_{l}^{\mathrm{op}}$, which leads to~\eqref{ru_du1} and~\eqref{wu_til}. This completes the proof. 

\section{Proof of Theorem~\ref{theo2}}\label{apd3}
According to the definition,
\begin{align}
   P_{\mathrm{A}}(\Omega, \bm{\eta}) &= |\alpha|^{2}\mathbf{v}^{\mathrm{H}}(\Omega)\operatorname{diag}\{\bm{\eta}^{\ast}\}\mathbf{V}\mathbf{V}^{\mathrm{H}}\operatorname{diag}\{\bm{\eta}\}\mathbf{v}(\Omega)\nonumber\\ 
   &= \frac{|\alpha|^{2}}{M^{2}}\sum_{m = 1}^{M}\eta^{\ast}_{m}\sum_{m^{\prime} = 1}^{M}\eta_{m^{\prime}}\Xi_{m,m^{\prime}}(\Omega),
\end{align}
where $\Xi_{m,m^{\prime}}(\Omega) \triangleq \sum_{a=1}^{A}\exp\{\overline{\jmath}2\pi f_{\mathrm{c}}\Delta_{\tau}(m - m^{\prime})(\Omega - \Omega_{a})\}$. Combining $\Omega_{a}= \left(a - 1 - \left\lfloor \Delta_{\mathrm{an}}^{-1}\right\rfloor\right)\Delta_{\mathrm{an}}$, $S = 1/(f_{\mathrm{c}}\Delta_{\tau}\Delta_{\mathrm{an}})$ and $A = 2\left\lfloor \Delta_{\mathrm{an}}^{-1}\right\rfloor + 1$ yields  
\begin{align}
   \Xi_{m,m^{\prime}}(\Omega) =& \exp\{\overline{\jmath}2\pi(m - m^{\prime})\Delta_{\mathrm{an}}^{-1}\Omega/S\}\nonumber\\
   & \times D_{(A-1)/2}(2\pi(m - m^{\prime})/S).
\end{align}
Furthermore,
\begin{align}
   \int_{-\Omega^{\prime}}^{\Omega^{\prime}}& P_{\mathrm{A}}(\Omega, \bm{\eta})\mathrm{d}\Omega \nonumber\\
   =& \frac{|\alpha|^{2}}{M^{2}}\sum_{m = 1}^{M}\eta^{\ast}_{m}\sum_{m^{\prime} = 1}^{M}\eta_{m^{\prime}}\int_{-\Omega^{\prime}}^{\Omega^{\prime}}\Xi_{m,m^{\prime}}(\Omega)\mathrm{d}\Omega\nonumber\\
   =& \frac{2\Omega^{\prime}|\alpha|^{2}}{M^{2}}\sum_{m = 1}^{M}\eta^{\ast}_{m}\sum_{m^{\prime} = 1}^{M}\eta_{m^{\prime}}\mathrm{sinc}(2\pi(m - m^{\prime})\Omega^{\prime}\Delta_{\mathrm{an}}^{-1}/S)\nonumber\\
   &\times D_{(A-1)/2}(2\pi(m - m^{\prime})/S).
\end{align}
On the other hand,
\begin{equation}
   P_{\mathrm{C}}(\Omega, \bm{\eta}) = \frac{|\alpha|^{2}}{M^{2}}\sum_{m = 1}^{M}\eta^{\ast}_{m}\sum_{m^{\prime} = 1}^{M}\eta_{m^{\prime}}\Phi_{m,m^{\prime}}(\Omega),
\end{equation}
where 
\begin{align}
   \Phi_{m,m^{\prime}}(\Omega) \triangleq & \sum_{a=a(\Omega)-c}^{a(\Omega)+c}\exp\{\overline{\jmath}2\pi f_{\mathrm{c}}\Delta_{\tau}(m - m^{\prime})(\Omega - \Omega_{a})\}\nonumber\\
   =& \exp\{\overline{\jmath}2\pi(m - m^{\prime})(\Delta_{\mathrm{an}}^{-1}\Omega - \lceil\Delta_{\mathrm{an}}^{-1}\Omega\rfloor)/S\}\nonumber\\
   &\times D_{c}(2\pi(m - m^{\prime})/S).
\end{align}
Therefore,
\begin{align}
   \int_{-\Omega^{\prime}}^{\Omega^{\prime}}& P_{\mathrm{C}}(\Omega, \bm{\eta})\mathrm{d}\Omega \nonumber\\
   =& \frac{|\alpha|^{2}}{M^{2}}\sum_{m = 1}^{M}\eta^{\ast}_{m}\sum_{m^{\prime} = 1}^{M}\eta_{m^{\prime}}\int_{-\Omega^{\prime}}^{\Omega^{\prime}}\Phi_{m,m^{\prime}}(\Omega)\mathrm{d}\Omega\nonumber\\
   =& \frac{2\Omega^{\prime}|\alpha|^{2}}{M^{2}}\sum_{m = 1}^{M}\eta^{\ast}_{m}\sum_{m^{\prime} = 1}^{M}\eta_{m^{\prime}}\mathrm{sinc}(\pi(m - m^{\prime})/S)\nonumber\\
   &\times D_{c}(2\pi(m - m^{\prime})/S).
\end{align}
As a result, 
\begin{equation}
   \lambda(\bm{\eta}) = \frac{\int_{-\Omega^{\prime}}^{\Omega^{\prime}}P_{\mathrm{C}}(\Omega, \bm{\eta})\mathrm{d}\Omega}{\int_{-\Omega^{\prime}}^{\Omega^{\prime}}P_{\mathrm{A}}(\Omega, \bm{\eta})\mathrm{d}\Omega} = \frac{\bm{\eta}^{\mathrm{H}}\mathbf{\Phi}\bm{\eta}}{\bm{\eta}^{\mathrm{H}}\mathbf{\Xi}\bm{\eta}}.
\end{equation}
Thus \eqref{opt_win} can be formulated as a maximization problem of the generalized Rayleigh quotient~\cite{9592715}, whose optimal solution is the generalized eigenvector associated with the maximum generalized eigenvalue of $\{\bm{\Phi}, \bm{\Xi}\}$. This completes the proof.

\section{Proof of Corollary~\ref{coro1}}\label{apd4}
Denote $\overline{\lambda}$ as the maximum generalized eigenvalue of $\{\bm{\Phi},\bm{\Xi}\}$, and $\overline{\bm{\eta}}$ is the corresponding generalized eigenvector. Set $\bm{\Psi} = \bm{\Phi} - \overline{\lambda}\bm{\Xi}$. According to the definition of generalized eigenvector, we have 
\begin{equation}\label{A_eta0}
   \bm{\Psi}\overline{\bm{\eta}} = \mathbf{0}_{M}.
\end{equation} 
From \eqref{Phi_Xi}, it can be seen that $\bm{\Phi}$ and $\bm{\Xi}$ are both real symmetric Toeplitz matrices, thus $\overline{\bm{\eta}}$ can be selected as a real vector. And $\bm{\Psi}$ is also real symmetric Toeplitz matrix, which satisfies $\bm{\Psi}\overline{\mathbf{I}}_{M} = \overline{\mathbf{I}}_{M}\bm{\Psi}$. Consequently,
\begin{equation}\label{A_eta_inv}
   \bm{\Psi}\overline{\mathbf{I}}_{M}\overline{\bm{\eta}} = \overline{\mathbf{I}}_{M}\bm{\Psi}\overline{\bm{\eta}} = \mathbf{0}_{M}.
\end{equation}
We next construct $\bm{\eta}^{\mathrm{o}} = \overline{\bm{\eta}} + \overline{\mathbf{I}}_{M}\overline{\bm{\eta}}\in\mathbb{R}^{M}$. From \eqref{A_eta0} and \eqref{A_eta_inv} we have $\bm{\Psi}\bm{\eta}^{\mathrm{o}} = \bm{\Psi}\overline{\bm{\eta}} + \bm{\Psi}\overline{\mathbf{I}}_{M}\overline{\bm{\eta}} = \mathbf{0}_{M}$. Therefore, $\bm{\eta}^{\mathrm{o}}$ is a generalized eigenvector corresponding to $\overline{\lambda}$, i.e., an optimal solution of \eqref{opt_win}. Additionally, $\overline{\mathbf{I}}_{M}\bm{\eta}^{\mathrm{o}} = \overline{\mathbf{I}}_{M}\overline{\bm{\eta}} + \overline{\bm{\eta}} = \bm{\eta}^{\mathrm{o}}$, which indicates $\bm{\eta}^{\mathrm{o}}$ exhibits real centrosymmetric structure. This completes the proof. 

\section{Proof of Lemma~\ref{lemm2}}\label{apd5}
Denote $\overline{\mathbf{Q}} = \mathbf{F}_{S}^{\mathrm{H}}\mathbf{I}_{S, M}\bm{\Lambda}\mathbf{I}_{M, S}\mathbf{F}_{S}$. Since $\mathbf{I}_{S, M}\bm{\Lambda}\mathbf{I}_{M, S} = \operatorname{diag}\{\mathbf{I}_{S,M}\bm{\eta}\}$ is a diagonal matrix, then $\overline{\mathbf{Q}}$ is a circulant matrix, which can be expressed as $\overline{\mathbf{Q}} = \sum_{k=0}^{S-1}\overline{\gamma}_{k}\bm{\Pi}_{S}^{k}$ with $\overline{\gamma}_{k} = \mathbf{f}_{S,k+1}^{\mathrm{H}}\mathbf{I}_{S,M}\bm{\eta}$. Due to the centrosymmetry $\bm{\eta} = \overline{\mathbf{I}}_{M}\bm{\eta}$, 
\begin{align}
   \overline{\gamma}_{k} &= \frac{1}{2}\mathbf{f}_{S,k+1}^{\mathrm{H}}\mathbf{I}_{S,M}(\mathbf{I}_{M} + \overline{\mathbf{I}}_{M})\bm{\eta} \nonumber\\
   &= \exp\{\overline{\jmath}\pi k(M-1)/S\}\mathbf{c}_{k}^{\mathrm{T}}\bm{\eta}.
\end{align}
Thus $\overline{\mathbf{Q}} = M\sum_{k=0}^{S-1}\exp\{\overline{\jmath}\pi k(M-1)/S\}\overline{\gamma}_{k}\bm{\Pi}_{S}^{k}$. On the other hand, $\bm{\Omega} = \operatorname{diag}\{\mathbf{I}_{S,2S}\mathbf{f}_{2S,M}^{\ast}\}$ is a diagonal matrix, according to the identity $\operatorname{diag}\{\mathbf{a}\}\mathbf{A}\operatorname{diag}\{\mathbf{b}\} = \mathbf{A}\odot (\mathbf{a}\mathbf{b}^{\mathrm{T}})$, we have 
\begin{align}
   \bm{\Omega}^{\ast}\overline{\mathbf{Q}}\bm{\Omega} =& \overline{\mathbf{Q}} \odot \big(\mathbf{I}_{S,2S}\mathbf{f}_{2S,M}\mathbf{f}_{2S,M}^{\mathrm{H}}\mathbf{I}_{S,2S}\big) \nonumber\\
   =& M\left(\sum_{k=0}^{S-1}\exp\{\overline{\jmath}\pi k(M-1)/S\}\overline{\gamma}_{k}\bm{\Pi}_{S}^{k}\right)\nonumber\\
   &\odot \left(\sum_{k^{\prime}=0}^{S-1}\exp\{-\overline{\jmath}\pi k(M-1)/S\}\overline{\bm{\Pi}}_{S}^{k^{\prime}}\right).
\end{align}
According to the definition of $\bm{\Pi}_{S}$ and $\overline{\bm{\Pi}}_{S}$, 
\begin{equation}
   \bm{\Pi}_{S}^{m}\odot \overline{\bm{\Pi}}_{S}^{n} =
   \left\{\begin{aligned}
      &\overline{\bm{\Pi}}_{S}^{n},&& \langle m-n\rangle_{S} = 0 \\
      &\mathbf{O},&& \langle m-n\rangle_{S} \neq 0
       \end{aligned}\right. .
\end{equation}
Then $\bm{\Omega}^{\ast}\overline{\mathbf{Q}}\bm{\Omega} = M\sum_{k=0}^{S-1}\gamma_{k}\overline{\bm{\Pi}}_{S}^{k}$. Followed by the property $\overline{\bm{\Pi}}_{S}^{(A-1)/2}\overline{\bm{\Pi}}_{S}^{k}\overline{\bm{\Pi}}_{S}^{S-(A-1)/2} = -\overline{\bm{\Pi}}_{S}^{k}$,
\begin{equation}
   \widetilde{\mathbf{Q}} = -\frac{1}{M}\overline{\bm{\Pi}}_{S}^{(A-1)/2}\bm{\Omega}^{\ast}\overline{\mathbf{Q}}\bm{\Omega}\overline{\bm{\Pi}}_{S}^{S-(A-1)/2} = \sum_{k=0}^{S-1}\gamma_{k}\overline{\bm{\Pi}}_{S}^{k}.
\end{equation}
Since $\gamma_{k}=\frac{1}{M}\mathbf{c}_{k}^{\mathrm{T}}\bm{\eta}\in\mathbb{R},\ k\in\mathbb{Z}$, and $\overline{\bm{\Pi}}_{S}^{k}\in\mathbb{R}^{S\times S},\ k\in\mathbb{Z}$, are real-valued Toeplitz matrices, then $\widetilde{\mathbf{Q}}$ is a real-valued Toeplitz matrix.
It can be checked that $\mathbf{c}_{k} = (-1)^{M+1}\mathbf{c}_{S-k},\ k\in\mathbb{Z}$, then we have $\gamma_{k} = (-1)^{M+1}\gamma_{S-k},\ k\in\mathbb{Z}$. Note that $\gamma_{0} = \frac{1}{M}\operatorname{tr}\{\bm{\Lambda}\} = 1$, then $\widetilde{\mathbf{Q}}$ can be expressed as
\begin{equation}
   \widetilde{\mathbf{Q}} = \mathbf{I}_{S} + \sum_{k=1}^{\lceil S/2\rceil-1}\gamma_{k}\left(\overline{\bm{\Pi}}_{S}^{k} + (-1)^{M+1}\overline{\bm{\Pi}}_{S}^{S-k}\right).
\end{equation}
This completes the proof.

\bibliographystyle{IEEEtran}
\bibliography{reference}

\end{document}